\documentclass[aip,jcp,reprint]{revtex4-2}
\usepackage[version=3]{mhchem} 
\usepackage{amsmath,amsthm,amssymb,amsfonts}
\usepackage{float}
\usepackage{siunitx}
\usepackage{graphicx}
\usepackage{booktabs}
\usepackage[english]{babel}
\usepackage{textgreek}
\usepackage{hyperref}
\usepackage[utf8]{inputenc}
\usepackage[T1]{fontenc}
\usepackage{mathptmx}
\usepackage{etoolbox}
\usepackage{dcolumn}
\usepackage{bm}
\usepackage{xcolor}

\usepackage{tikz}
\usetikzlibrary{arrows.meta, positioning, fit, backgrounds, calc, shapes, shapes.geometric}

\makeatletter
\def\@email#1#2{%
 \endgroup
 \patchcmd{\titleblock@produce}
  {\frontmatter@RRAPformat}
  {\frontmatter@RRAPformat{\produce@RRAP{*#1\href{mailto:#2}{#2}}}\frontmatter@RRAPformat}
  {}{}
}%
\makeatother

\DeclareSIUnit\angstrom{\text {Å}}
\begin{document}


\title{Investigating causality between principal components in protein dynamics}

\author{Debarshi Banerjee}
\affiliation{Scuola Internazionale Superiore di Studi Avanzati (SISSA), Via Bonomea 265, 34136 Trieste, Italy}
\affiliation{International Centre for Theoretical Physics (ICTP), Strada Costiera 11, 34151 Trieste, Italy}

\author{Ali Hassanali}
\email{ahassana@ictp.it}
\affiliation{International Centre for Theoretical Physics (ICTP), Strada Costiera 11, 34151 Trieste, Italy}

\author{Alessandro Laio}
\email{laio@sissa.it}
\affiliation{Scuola Internazionale Superiore di Studi Avanzati (SISSA), Via Bonomea 265, 34136 Trieste, Italy}
\affiliation{International Centre for Theoretical Physics (ICTP), Strada Costiera 11, 34151 Trieste, Italy}




\begin{abstract}
Principal component analysis (PCA) is widely used to characterize collective protein motions from molecular dynamics (MD) simulations. While PCA identifies the dominant modes of structural fluctuation, it does not reveal whether different principal components (PCs) causally influence each other. Here, we investigate this question using a recently introduced causal-discovery framework [Del Tatto et al, PNAS 2024], which allows to infer putative causal asymmetries between high-dimensional time series.
We apply this approach to long-timescale MD trajectories of two proteins. By analyzing  relationships among PCs, we construct directed networks describing how PCs influence one another across time scales. 
These directional relationships, whose existence is a necessary condition for the presence of a causal link,  are not captured by conventional covariance-based analyses and provide information that is complementary to PCA and Time-lagged Independent Component Analysis (TICA). 
Our results suggest that our causal inference approach can uncover previously hidden aspects of the dynamical organization of protein motions and offer a new perspective on this very popular class of collective variables.
\end{abstract}

\maketitle


\section{Introduction} \label{sec:intro}

Understanding how collective motions in proteins are organized across multiple time scales remains a central challenge in molecular biophysics\cite{karplus2002molecular,henzler2007dynamic,boehr2009role,guo2016protein,nam2023protein,grandori2023protein,ono2023conformational,haran2024fast,hatton2024exploring}. Molecular dynamics (MD) simulations now routinely generate trajectories of sufficient length and resolution to characterize fluctuations ranging from fast local motions to rare conformational transitions\cite{shaw2009millisecond,klepeis2009long,DEShaw2011,dror2012biomolecular,lane2013milliseconds,hollingsworth2018molecular,huggins2019biomolecular,shaw2021anton,bhati2023long}, enabling the investigation of both thermodynamic and kinetic properties. In particular, equilibrium trajectories of small proteins extending to biologically relevant time scales are now available\cite{shaw2010atomic,DEShaw2011,DEShaw2016}, providing unprecedented opportunities to study the relationship between structure, dynamics, and function.

One of the most widely used approaches for analyzing protein dynamics in the folded state is Principal Component Analysis (PCA)\cite{kitao1991effects,ichiye1991collective,garcia1992large,amadei1993essential,hayward1995harmonicity,hayward1995collective,balsera1996principal,kitao1999investigating,maisuradze2009principal,david2013principal,cossio2017consistent,kitao2022principal}. In its standard formulation, PCA is performed by computing the covariance matrix of the positions of a selected subset of protein atoms (for example, backbone \(\alpha\) carbon atoms) after removing overall translational and rotational motions. The principal components (PCs) are the eigenvectors of this covariance matrix, while the corresponding eigenvalues quantify the variance along each component.\cite{buchner1992short,hayward1995collective,kitao1999investigating,horstink1999functionally,berendsen2000collective,david2013principal,kitao2022principal} When only a small number of eigenvalues account for a large fraction of the total variance, protein motions can be represented in a reduced-dimensional basis spanned by the leading PCs. In this basis, the covariance matrix is diagonal by construction, meaning that fluctuations along different PCs are uncorrelated at zero time lag. Owing to its ability to provide a compact description of collective structural fluctuations and to identify large-amplitude motions, PCA has become a standard tool for characterizing conformational landscapes and functionally relevant motions in proteins\cite{chau1999functional,horstink1999functionally,mu2005energy,capozzi2007essential,altis2007dihedral,pontiggia2008small,altis2008construction,maisuradze2009principal,morra2012corresponding,maisuradze2013local,david2013principal,sittel2014principal,ernst2015contact,post2019principal,kitao2022principal}.

When the focus shifts from structural characterization to dynamical behavior, Time-lagged Independent Component Analysis (TICA) provides a natural extension of PCA.\cite{molgedey1994TICA,naritomi2011TICA,perez2013TICA,schwantes2013TICA} While principal components are uncorrelated at zero lag time, correlations generally arise between them when the system is observed at a finite time delay, $\tau$. $C(\tau)$ denotes the time-lagged covariance matrix between the centered and aligned atomic coordinates at times 0 and $\tau$. In the PCA basis, the corresponding lagged covariance matrix is given by $\Gamma(\tau)=C^{-1}(0)C(\tau)$.\cite{molgedey1994TICA,perez2013TICA,noe2015kinetic,perez2016hierarchical} This matrix would remain diagonal for all values of $\tau$ only in the idealized case of purely harmonic dynamics. For realistic protein motions $\Gamma(\tau)$ is generally non-diagonal, reflecting dynamical couplings between different PCs. The eigenvectors of $\Gamma(\tau)$, commonly referred to as time-lagged independent components (ICs), identify collective coordinates associated with the slowest dynamical processes. As a result, they are widely employed to characterize long-timescale conformational changes, construct kinetic models, and define collective variables for enhanced-sampling simulations\cite{perez2013TICA,schwantes2013TICA,nuske2014variational,noe2015kinetic,sultan2017tica,schultze2021time,bonati2021deep}.

These developments raise a natural question: can collective structural fluctuations identified by PCA exert asymmetric dynamical influences on one another? More specifically, can principal components extracted from protein MD trajectories \emph{cause} each other?\cite{pearl2009causality,spirtes2000causation,runge2018causal}  In the language of causal inference, stating that, for example, \(\mathrm{PC}_2\) is a putative cause of \(\mathrm{PC}_4\) means that knowledge of the present state of \(\mathrm{PC}_2\) improves the prediction of the future evolution of \(\mathrm{PC}_4\) beyond what can be inferred from the history of \(\mathrm{PC}_4\) alone\cite{granger1969investigating}. A natural first attempt to address this question would be to examine the time-lagged covariance matrix between PCs, $\Gamma(\tau)$. If, for instance, the off-diagonal element $\Gamma_{2\,4}(\tau)$ is large, one may conclude that \(\mathrm{PC}_2\) and \(\mathrm{PC}_4\) are dynamically coupled. However, dynamical coupling does not necessarily imply causal influence. In particular, causal relationships are generally directional: it may happen that \(\mathrm{PC}_2\) influences the future evolution of \(\mathrm{PC}_4\), while the converse is not true. 
Such asymmetries, if present, cannot be inferred from $\Gamma(\tau)$ alone. For reversible equilibrium dynamics, the transfer operator is self-adjoint with respect to the equilibrium measure. Consequently, TICA identifies slow decorrelating coordinates but cannot assign a directional driver--response interpretation to couplings between pairs of PCs.

Addressing this question requires tools that go beyond covariance or cross-correlation based analyses. 
Several approaches have been proposed to infer causality from molecular simulations. Early studies employed multivariate autoregressive models and analyses inspired by Granger causality to identify directed relationships between coordinates or residue fluctuations.\cite{gorecki2006causal,kamberaj2009extracting,qi2013quantification,sobieraj2022granger} An information-theoretic approach, dubbed transfer entropy\cite{schreiber2000measuring,schindler2007causality}, has also been used to probe directional information transfer and allosteric communication in proteins\cite{hacisuleyman2017entropy,hacisuleyman2017causality,hempel2020coupling}. 
More broadly, causal inference for time series has developed into  rich theoretical frameworks aimed at predicting information asymmetries, conditional dependencies, and structural causal graphs\cite{granger1969investigating,pearl2009causality,runge2018causal,runge2023causal_inference}. In practice, however, applying these approaches to atomistic MD remains challenging because the statistical errors of causal estimators are typically large when the causal links are weak.

To address these difficulties, we recently introduced the Imbalance Gain (IG), a statistical test designed to identify putative causal  couplings in time-dependent data\cite{glielmo2022ranking,deltatto2024robust,DelTatto2025}. The IG can be viewed as a proxy for the transfer entropy\cite{schreiber2000measuring,schindler2007causality}  and can also be efficiently estimated for noisy and high-dimensional data.\cite{deltatto2024robust} 
It is also related to Granger causality and related methods but, unlike those approaches, it  does not require fitting an explicit dynamical model. Recent studies have shown that putative causal couplings identified  through the IG can robustly reveal genuine causal relationships even in equilibrium molecular dynamics, despite the time-reversible nature of the microscopic equations of motion and the stationarity of the underlying distribution\cite{DelTatto2025}. The same framework has also uncovered  non-trivial  causal relationships involving translational and rotational degrees of freedom  in liquid water\cite{Huet2026}. It is important to note that the IG and other similar approaches infer the presence of \emph{putative} causal links from observational data. The presence of an  asymmetric dynamical coupling between two variables is a \emph{necessary} condition of the presence of a causal link between the two variables. However, this condition is not sufficient, since the two variables could be simultaneously caused by a third unobserved variable.  Revealing the presence of confounders and building a genuine causal graph\cite{pearl2009causality,runge2018causal,runge2023causal_inference} in molecular modeling remains an open challenge. 

In the present work, we use the IG framework to investigate whether the principal components describing collective protein motions exhibit  putative causal links. We analyze two long-timescale trajectories generated by D. E. Shaw Research: ubiquitin (PDB ID: 1UBQ)\cite{DEShaw2016} and NTL9 (PDB ID: 2HBA)\cite{DEShaw2011}.  NTL9 is a fast-folding protein whose trajectory samples large conformational changes and multiple folding and unfolding events\cite{DEShaw2011}, whereas ubiquitin represents a prototypical native-state system whose equilibrium fluctuations span timescales from picoseconds to milliseconds and involve subtle long-range collective motions\cite{DEShaw2016}. By comparing these two systems, we assess whether putative causal links among PCs emerge, and whether such links are controlled by variance, relaxation timescale, or spatial collectivity. We find that both proteins exhibit a causal structure, but with distinct physical character: in ubiquitin, the directionality of the putative causal link originates from a highly collective, protein-wide mode and affects only the next few leading components, whereas in NTL9, a localized slow mode centered on the \(\beta\)-turn directionally couples to several faster modes, including more delocalized principal components.

\begin{figure*}[ht!]
  \centering
  \includegraphics[width=1.0\linewidth]{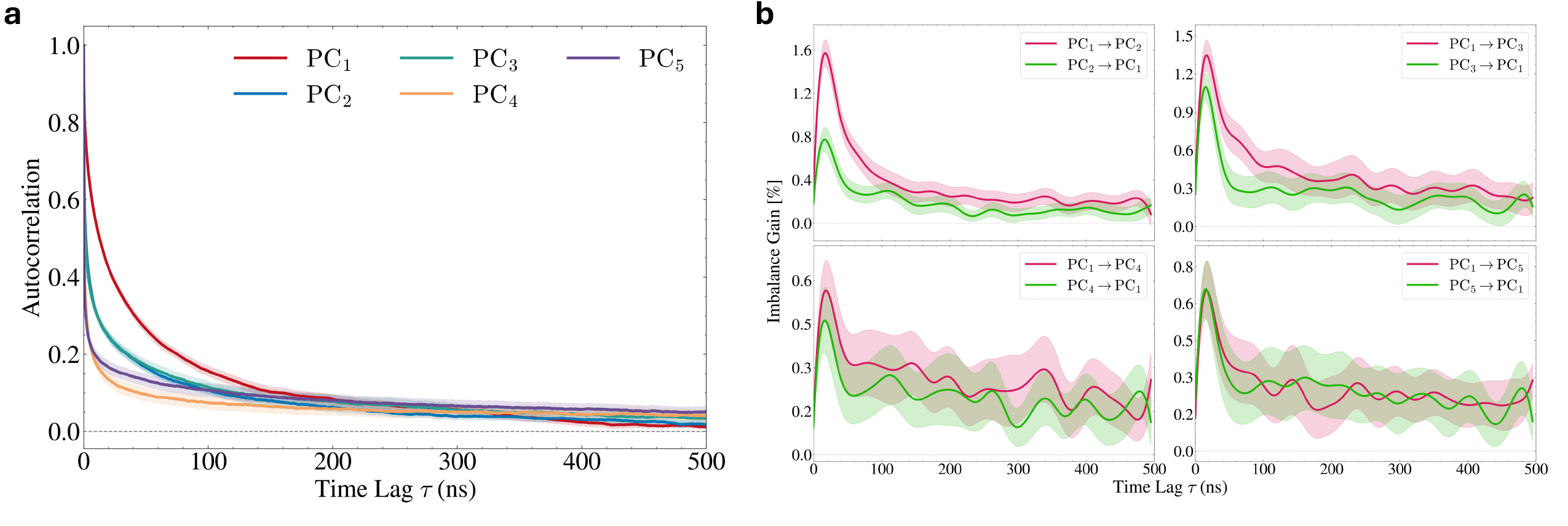}
  \caption{Ubiquitin:
(a) The autocorrelation functions of the top 5 PCs, $\mathrm{PC}_{i=1,5}$, which are colored red, blue, green, orange, and purple respectively, with shaded areas denoting error bars.
(b) The pairwise IG curves over time between $\mathrm{PC}_1 \Leftrightarrow \mathrm{PC}_{2,3,4,5}$. The IG values are on the Y-axis, and time lag ($\tau$) in nanoseconds is on the X-axis. Pink lines denote the direction $\mathrm{PC}_1 \rightarrow \mathrm{PC}_{2,3,4,5}$, whereas green lines denote the reverse direction. Shaded areas denote error bars computed over 20 independent realizations. 
}
  \label{fig:ubq_acf_ig}
\end{figure*}

\section{Methods} \label{sec:methods}


Within the Imbalance  Gain (IG) framework\cite{deltatto2024robust}, a putative causal link between two (possibly multidimensional) collective variables $X$ and $Y$ is estimated in terms of predictive asymmetry between the two  time series $X(t)$ and $Y(t)$. The IG is based on the Information Imbalance\cite{glielmo2022ranking} (II), a statistical framework that provides a nonparametric way to compare the information content of different representations of the same data by exploiting the relative ordering of distances in the two spaces, $X$ and $Y$. In particular, the II quantifies how well neighborhoods defined in $X$ are preserved in $Y$. Denoting by $r_{ij}^X$ the distance rank of configuration $j$ among the neighbors of $i$ in space $X$, the II of $X$ with respect to $Y$ is defined as
\begin{equation}\label{eq:iib}
\Delta(X \to Y)
=
\frac{2}{N}
\left\langle r^{Y}_{ij} \,\middle|\, r^{X}_{ij} \le k \right\rangle = \frac{2}{N^2\,k} \sum_{i,j:\, r^{ij}_X\, \leq k} r^{ij}_Y\,
\end{equation}
where $N$ is the total number of configurations and $k$ is a small number (here set to 50, which is about 2\% of $N$) used to  restrict the average to the closest neighbors in space $X$. $\Delta(X \to Y)$ yields a value between 0 and 1. Small values of $\Delta(X \to Y)$ indicate that neighborhoods in $X$ are well preserved in $Y$, implying that $X$ contains most of the information needed to reconstruct $Y$. Values closer to 1 instead indicate that $X$ is minimally informative with respect to $Y$. The II framework has been used to infer non-trivial relationships between collective variables in a wide variety of applications in molecular systems.\cite{Donkor2023,Donkor2024,DiPino2025,banerjee2026machine}

The presence of a putative causal link is then inferred by estimating the II between collective variables at different times. Qualitatively, one estimates whether the inclusion of a candidate driver variable improves the prediction of a target variable at a future time.\cite{deltatto2024robust,DelTatto2025} Since the Information Imbalance is computed over a set of points, we suppose to have access to multiple replicas of the trajectory. If the available data consist of a single trajectory, an ensemble of replicas is constructed by dividing the trajectory into 
 non-overlapping  chunks. For a target variable $Y(0)$ and a candidate driver $X(0)$, one compares the II between the present state of the target and its future, $\Delta\!\big(Y(0) \to Y(\tau)\big)$, with the II obtained when also considering the putative driver in the present, $\Delta\!\big([\alpha X(0),Y(0)] \to Y(\tau)\big)$,
where $\alpha > 0$ is a variational parameter. 
In these equations $Y(0)$,  $X(0)$, and $[\alpha X(0),Y(0)]$  are a shorthand notation for the Euclidean distance between pairs of replicas. For example, $[\alpha X(0),Y(0)] = (\alpha^2 ||X_i(0) - X_j(0)||^2 + ||Y_i(0) - Y_j(0)||^2)^\frac{1}{2}$, where the subscripts $i$ and $j$ label different replicas.

The Imbalance Gain (IG) is then defined as
\begin{equation}\label{eq:ig}
\mathrm{IG}_{X \to Y}(\tau) 
= 
\frac{ \Delta\!\big(Y(0) \to Y(\tau)\big)- 
\min\limits_{\alpha}\Delta\!\left([\alpha X(0),\,Y(0)] \to Y(\tau)\right)}{\Delta\!\big(Y(0) \to Y(\tau)\big)}
\end{equation}
The optimal value of $\alpha$ that maximizes the IG for a given time lag $\tau$ is determined by searching over a grid of 500 evenly spaced values for $\alpha \in [0,10]$.
If IG=0, adding the information about $X(0)$ does not improve the prediction of $Y(\tau)$, since the best possible predictor is $Y(0)$ alone. 
A positive value of the $\mathrm{IG}_{X \to Y}(\tau)$ indicates instead that the inclusion of $X(0)$ improves the prediction of the future state of $Y$, providing evidence for a putative causal link at lag $\tau$. As already discussed, the observation of  IG>0  between two variables does not imply that a causal link between the two exists, since both variables could be caused by a third unobserved variable. However, if one finds that IG=0 (or small, in practical applications) one can conclude that the causal link does not exist\cite{deltatto2024robust}. Since the IG is estimated using  distances between configurations, it naturally accommodates high-dimensional descriptors and avoids the computational problems that are encountered when one attempts to infer the causal links from transfer entropy, which requires estimating conditional probability distributions in high-dimensional spaces.\cite{schreiber2000measuring,schindler2007causality}

We also quantify a putative causal link between  two variables from the integral of the IG over time (IG Total):
\begin{equation}\label{eq:igt}
\mathrm{IGT}_{X \to Y}
=
\int_{0}^{\infty} \mathrm{IG}_{X \to Y}(\tau') d\tau'
\end{equation}

In the present work, we apply this framework to PCs obtained from protein dynamics. Treating individual PCs as time-dependent variables, we evaluate the lag-dependent IG between pairs of components to assess whether specific modes systematically improve the prediction of others. This allows us to infer the presence of putative causal links in principal component space. To simplify the visualization and the discussion, we restrict our analysis to the top five PCs.

\begin{figure*}[ht!]
  \centering
  \includegraphics[width=1.0\linewidth]{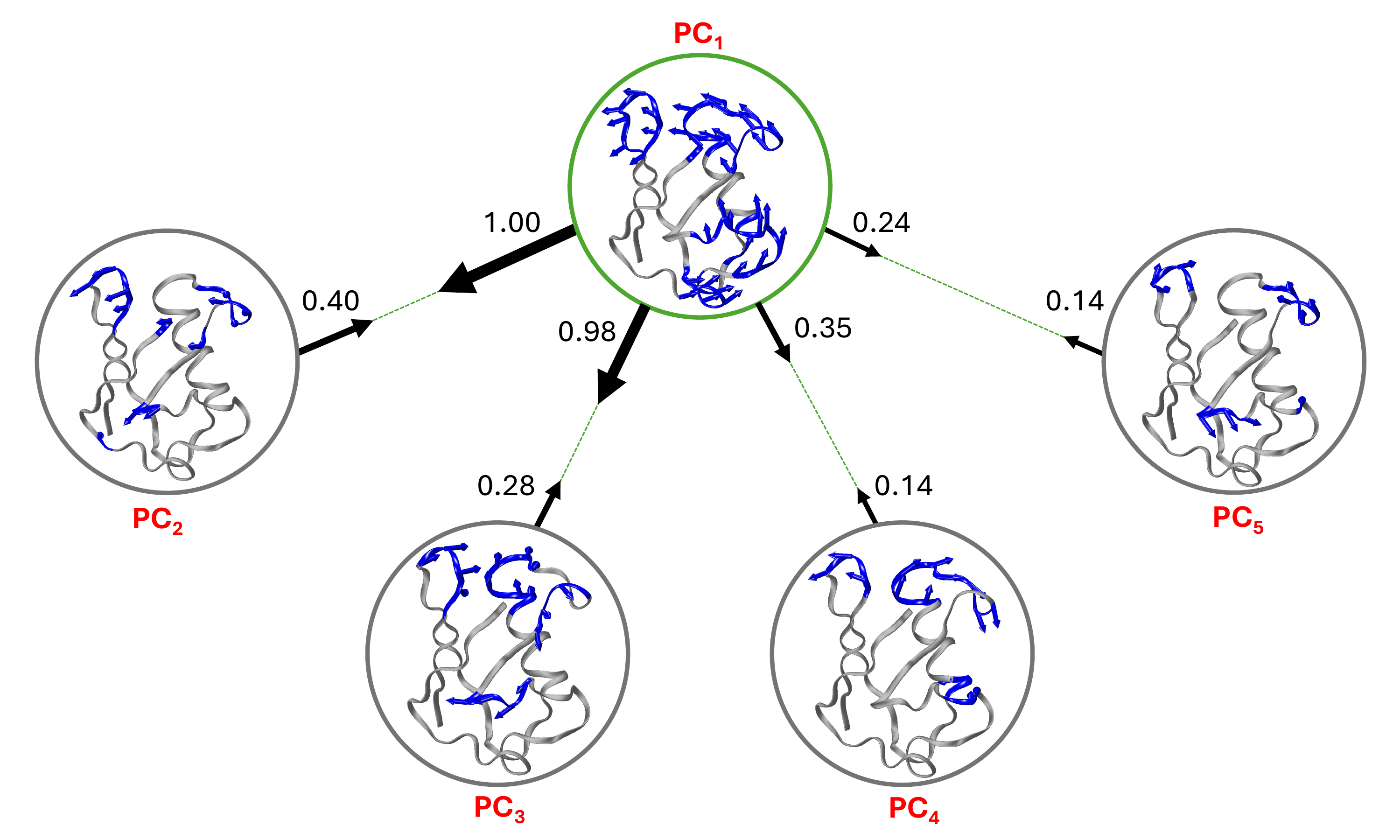}
  \caption{Ubiquitin: Putative causal links between \(\mathrm{PC}_1\) and the remaining leading principal components. Nodes represent the first five PCs, and directed edges indicate the normalized integrated Imbalance Gain, \(\mathrm{IGT}\), defined in Eq.~\ref{eq:igt}. 
For each pair, the edge length and width, as well as the numerical label report the relative strength of the directional information transfer, with larger values corresponding to stronger causal links. 
The strongest links are observed from \(\mathrm{PC}_1\) toward \(\mathrm{PC}_2\) and \(\mathrm{PC}_3\), whereas the links involving \(\mathrm{PC}_4\) and \(\mathrm{PC}_5\) are substantially weaker. 
For each PC, the residues with the largest contributions to the corresponding eigenvector are highlighted in blue on the protein structure, as quantified using the Br\"uschweiler collectivity index. 
The arrows in blue show the corresponding directions of the eigenvector for the highlighted residues, while the rest of the protein is shown in gray.
}
  \label{fig:ubq_pc1_graph}
\end{figure*}

\section{Results} \label{sec:results}

\subsection{Ubiquitin}

In the first application, we analyze the native-state dynamics of ubiquitin using the \(1~\mathrm{ms}\) molecular dynamics trajectory reported in Ref.~\citenum{DEShaw2016}. As discussed in that work, the trajectory exhibits a mild conformational transition at approximately \(0.65~\mathrm{ms}\). Since our goal is to characterize causal relationships between principal components within a single native-state basin, we restricted the analysis to the first \(0.65~\mathrm{ms}\) of the simulation. This choice avoids mixing fluctuations associated with distinct conformational substates and ensures that the resulting PCs describe a well-defined region of the native-state ensemble. The eigenvalue spectrum and the explained variance of the PCs are shown in SI Fig.~\ref{fig_SI:ubq_pca_spectrum}.

We first characterize the intrinsic relaxation timescales of the leading modes. In Fig.~\ref{fig:ubq_acf_ig}a, we report the autocorrelation functions of the first five principal components. All five modes decorrelate within approximately \(500~\mathrm{ns}\), which is substantially shorter than the analyzed trajectory segment. 
We determine the relaxation times from the integral of the autocorrelation function.
This indicates that the selected portion of the simulation is sufficiently long to sample the dominant fluctuations associated with these PCs. Among the modes considered, \(\mathrm{PC}_1\) is the slowest, while \(\mathrm{PC}_2\) and \(\mathrm{PC}_3\) decorrelate on comparable, shorter timescales. 

Previous studies on equilibrium MD simulations have used apparent asymmetries in time-lagged two-body cross-correlation functions, \(\mathbb{E}[ X_0 Y_\tau]\), to infer causal relationships \cite{dutta2017spatiotemporal,hacisuleyman2017entropy}. However, for equilibrium dynamics satisfying stationarity and time reversibility (as is the case for MD simulations), these correlation functions are invariant under exchange of the two variables. In particular,
\(
\mathbb{E} [X_0 Y_\tau]
=
\mathbb{E} [X_{-\tau}Y_0]
=
\mathbb{E} [X_\tau Y_0] ,
\)
where the first equality follows from stationarity and the second from time-reversibility. Therefore, any observed asymmetry in such correlation functions cannot be interpreted as evidence of directional causality. Instead, it must arise from statistical errors, numerical violations of time reversibility introduced by the integration scheme, or from perturbations associated with thermostats and barostats. The IG (or transfer entropy) approach is instead asymmetric and directional, and can be viewed conceptually as a many-body cross-correlation function where the information of the caused variable at the present time breaks the symmetry otherwise seen in two-body cross-correlation functions.\cite{deltatto2024robust,DelTatto2025}

\begin{figure*}[ht!]
  \centering
  \includegraphics[width=1.0\linewidth]{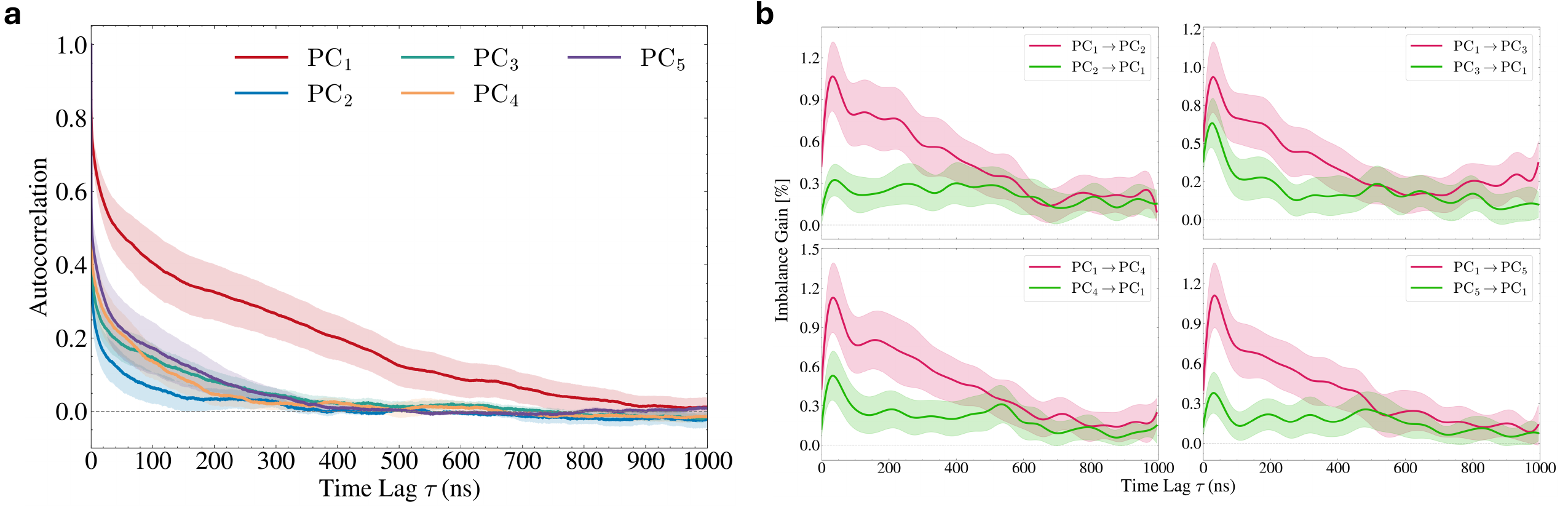}
  \caption{NTL9:
(a) The autocorrelation functions of the top 5 PCs, $\mathrm{PC}_{i=1,5}$, which are colored red, blue, green, orange, and purple respectively, with shaded areas denoting error bars.
(b) The pairwise IG curves over time between $\mathrm{PC}_1 \Leftrightarrow \mathrm{PC}_{2,3,4,5}$. The IG values are on the Y-axis, and time lag ($\tau$) in nanoseconds is on the X-axis. Pink lines denote the direction $\mathrm{PC}_1 \rightarrow \mathrm{PC}_{2,3,4,5}$, whereas green lines denote the reverse direction. Shaded areas denote error bars computed over 20 independent realizations. 
  }
  \label{fig:ntl9_acf_ig}
\end{figure*}

Therefore, we next examine putative causal links between the leading modes using the Imbalance Gain, instead of using two-body cross-correlation functions between the PCs. In Fig.~\ref{fig:ubq_acf_ig}b, we show the pairwise IG curves between \(\mathrm{PC}_1\) and \(\mathrm{PC}_{2,3,4,5}\). The strongest asymmetry is observed for the pairs \(\mathrm{PC}_1 \Leftrightarrow \mathrm{PC}_2\) and \(\mathrm{PC}_1 \Leftrightarrow \mathrm{PC}_3\), where the dominant direction of information transfer is \(\mathrm{PC}_1 \rightarrow \mathrm{PC}_{2,3}\). This indicates that knowledge of \(\mathrm{PC}_1\) improves the prediction of the future evolution of \(\mathrm{PC}_2\) and \(\mathrm{PC}_3\) more than the reverse. The asymmetry becomes weaker for \(\mathrm{PC}_4\), where the driving effect from \(\mathrm{PC}_1\) remains detectable but is considerably more modest. For \(\mathrm{PC}_5\), the IG signal is substantially reduced and becomes only weakly discernible. Consistently, when lower-variance PCs are considered, the directional signal progressively decreases and eventually approaches zero, as shown in the additional pairwise IG curves reported in the Supporting Information (see SI Fig.~\ref{fig_SI:ubq_pc1_rest_ig}). 
Overall, these results suggest that the dynamical influence of \(\mathrm{PC}_1\) is primarily exerted on the other dominant modes and progressively fades for less important PCs. This behavior is consistent with our empirical observations from both Langevin-based models as well as simple molecular systems, in which the slowest collective variable typically appears as a putative causal driver\cite{deltatto2024robust,DelTatto2025}.

The putative causal links between pairs of PCs are summarized in Fig.~\ref{fig:ubq_pc1_graph} using  the normalized IGT values defined in Eq.~\ref{eq:igt}. The graph highlights the dominant links between \(\mathrm{PC}_1\) and \(\mathrm{PC}_{2,3,4,5}\). In agreement with the time-resolved IG curves, the strongest directed links correspond to \(\mathrm{PC}_1 \rightarrow \mathrm{PC}_2\) and \(\mathrm{PC}_1 \rightarrow \mathrm{PC}_3\), confirming that the leading mode acts as the main driver of the next two dominant modes. A weaker directional asymmetry is also observed for the \(\mathrm{PC}_1 \Leftrightarrow \mathrm{PC}_{4,5}\) links, which remain biased toward the \(\mathrm{PC}_1 \rightarrow \mathrm{PC}_{4,5}\) direction. However, the magnitude of these links is markedly smaller than for \(\mathrm{PC}_2\) and \(\mathrm{PC}_3\), indicating that the causal influence of \(\mathrm{PC}_1\) rapidly weakens as one moves toward lower-variance PCs. As we will see in the next section, however, this is not necessarily a universal rule. 
In addition, SI Fig.~\ref{fig_SI:ubq_pc2_graph} shows that \(\mathrm{PC}_2\) exhibits appreciable asymmetric IG with both \(\mathrm{PC}_3\) and \(\mathrm{PC}_4\). In these cases, the dominant direction of information transfer is \(\mathrm{PC}_2 \rightarrow \mathrm{PC}_{3,4}\), while the corresponding reverse links are noticeably weaker. By contrast, the  IG between \(\mathrm{PC}_2\) and \(\mathrm{PC}_5\) is much smaller and only weakly directional. The same trend persists for lower-variance PCs, for which the causal signal rapidly decreases and eventually becomes negligible.

In Fig.~\ref{fig:ubq_pc1_graph}, we also highlight the residues that contribute most strongly to each principal component. These contributions were quantified using the Br\"uschweiler collectivity index, which quantifies the degree to which the collective motions are delocalized in the protein.\cite{Bruschweiler1995,Bruschweiler2000} Details are provided in SI Fig.~\ref{fig_SI:ubq_bruschweiller}. This analysis shows that \(\mathrm{PC}_1\) is the most delocalized mode among those considered, with a collectivity index of approximately \(0.8\), indicating that residues distributed across the entire protein contribute substantially to this collective fluctuation. By contrast, \(\mathrm{PC}_2\)--\(\mathrm{PC}_5\) are more localized, with collectivity indices in the range \(0.4\)--\(0.6\). The dominant role of \(\mathrm{PC}_1\) would suggest a scenario in which   more collective protein motions have a causal influence on more spatially localized fluctuations. In the following section we will show that this scenario does not hold in general.

\begin{figure*}[ht!]
  \centering
  \includegraphics[width=1.0\linewidth]{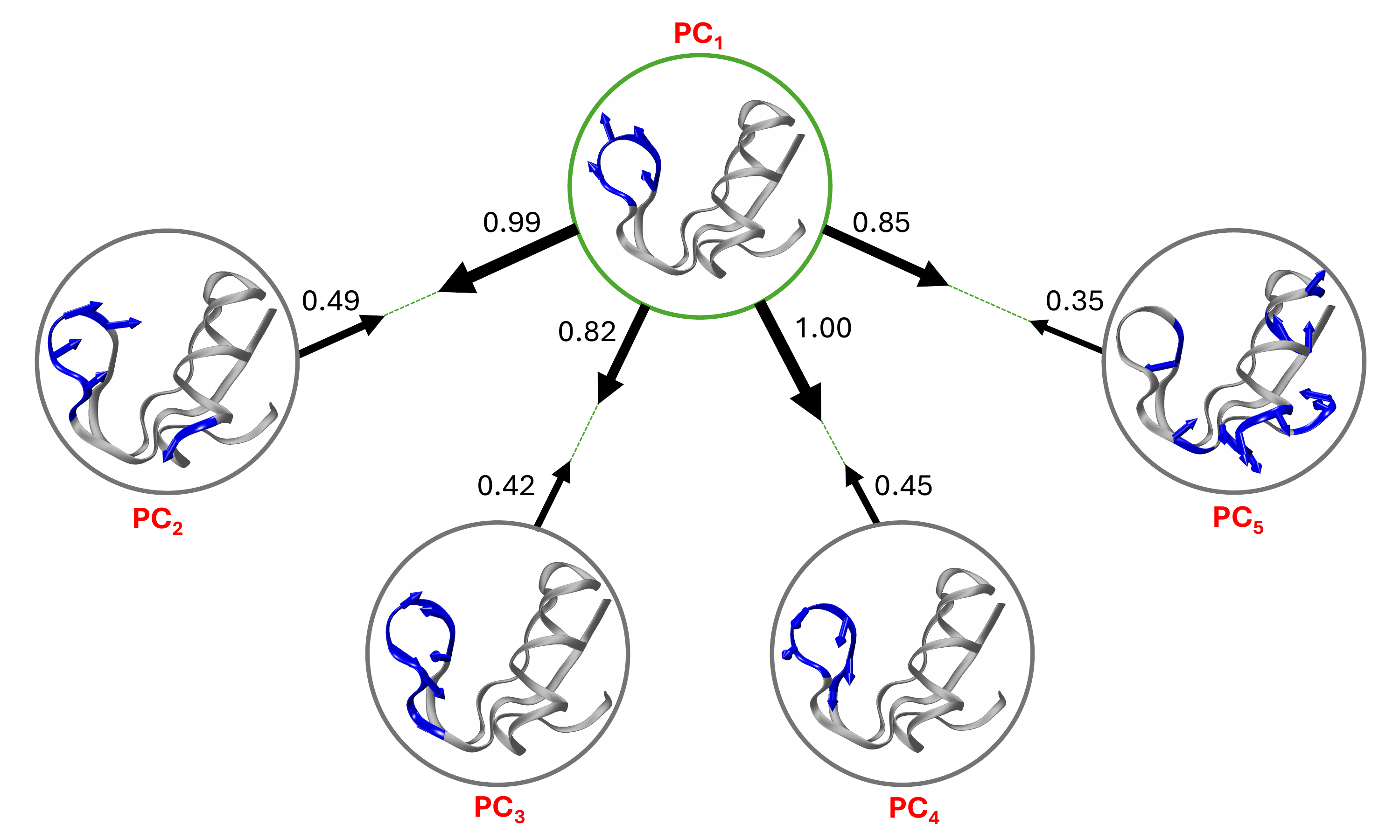}
  \caption{NTL9:
Putative causal links between \(\mathrm{PC}_1\) and the remaining leading principal components. 
Nodes represent the first five PCs, and directed edges indicate the normalized integrated Imbalance Gain, \(\mathrm{IGT}\), defined in Eq.~\ref{eq:igt}. 
For each pair, the edge length and width, as well as the numerical label report the relative strength of the directional information transfer, with larger values corresponding to stronger causal links. 
\(\mathrm{PC}_1\) displays strong directional influence toward all other leading PCs considered, identifying it as a dominant slow driver within the folded-state dynamics of NTL9. 
For each PC, the residues with the largest contributions to the corresponding eigenvector are highlighted in blue on the protein structure, as quantified using the Br\"uschweiler collectivity index.  
The arrows in blue show the corresponding directions of the eigenvector for the highlighted residues, while the rest of the protein is shown in gray.
}
\label{fig:ntl9_pc1_graph}
\end{figure*}

\subsection{NTL9}

In the second application, we analyze folded-state segments of the NTL9 trajectory. In Ref.~\citenum{DEShaw2011}, NTL9 was simulated for approximately \(3~\mathrm{ms}\), during which the protein undergoes multiple reversible folding--unfolding transitions. Within these simulations, the protein remains folded for several extended intervals; in particular, we identified four independent folded segments longer than \(200~\mu\mathrm{s}\). The results discussed in the following are obtained by analyzing one representative folded segment. The same analysis performed on the remaining folded segments yields qualitatively consistent results, as reported in the Supporting Information (see SI section ``NTL9 results for another folded trajectory fragment'' and SI Figs.~\ref{fig_SI:ntl9-0_acf} -- \ref{fig_SI:ntl9-0_bruschweiller}). This allows us to focus specifically on causal relationships between PCs within the folded-state dynamics, rather than on the much larger structural changes associated with folding and unfolding. The eigenvalue spectrum and the explained variance of the PCs are shown in SI Fig.~\ref{fig_SI:ntl9_pca_spectrum}.

Figure~\ref{fig:ntl9_acf_ig}a shows the autocorrelation functions of the first five principal components. In contrast to ubiquitin, where the leading modes decorrelate on more comparable sub-microsecond timescales, NTL9 displays a clearer separation between the slowest mode and the remaining PCs. In particular, \(\mathrm{PC}_1\) decorrelates on a timescale of approximately \(1~\mu\mathrm{s}\), whereas \(\mathrm{PC}_{2,3,4,5}\) relax substantially faster, with characteristic timescales of roughly \(200~\mathrm{ns}\). This separation indicates that \(\mathrm{PC}_1\) captures a distinct slow fluctuation within the folded-state ensemble of NTL9, while the remaining leading PCs describe faster modes embedded within the same folded basin.

We then examine whether this separation of timescales is reflected in the directionality of information transfer between the leading PCs. In Fig.~\ref{fig:ntl9_acf_ig}b, we report the pairwise IG curves between \(\mathrm{PC}_1\) and \(\mathrm{PC}_{2,3,4,5}\). In all cases, the curves show a pronounced asymmetry, with the \(\mathrm{PC}_1 \rightarrow \mathrm{PC}_{2,3,4,5}\) direction consistently dominating over the reverse direction. Thus, knowledge of the slow coordinate \(\mathrm{PC}_1\) substantially improves the prediction of the future evolution of the faster leading PCs, whereas the converse effect is much weaker. Compared with ubiquitin, where the directional influence of \(\mathrm{PC}_1\) is strongest for \(\mathrm{PC}_2\) and \(\mathrm{PC}_3\) but rapidly decreases for lower-variance modes, the NTL9 folded segment exhibits a more systematic and persistent directional coupling from \(\mathrm{PC}_1\) toward all the other  PCs considered (see SI Fig.~\ref{fig_SI:ntl9_pc1_rest_ig}).

This trend is summarized in Fig.~\ref{fig:ntl9_pc1_graph}, where we report the putative causal links constructed from the normalized IGT values defined in Eq.~\ref{eq:igt}. The graph confirms the behavior observed in the time-resolved IG curves: all links involving \(\mathrm{PC}_1\) are strongly asymmetric and predominantly oriented from \(\mathrm{PC}_1\) toward \(\mathrm{PC}_{2,3,4,5}\). By contrast, when \(\mathrm{PC}_2\) is taken as the reference mode, the links with \(\mathrm{PC}_{3,4,5}\) are largely reciprocal and of comparable magnitude, as shown in SI Fig.~\ref{fig_SI:ntl9_pc2_graph}. Taken together, these results identify \(\mathrm{PC}_1\) as a dominant slow mode that directionally conditions the dynamics of several faster principal components within the folded state of NTL9.

We again use the Br\"uschweiler collectivity index\cite{Bruschweiler1995,Bruschweiler2000} to identify the residues that contribute most strongly to each principal component; these residues are highlighted in blue in Fig.~\ref{fig:ntl9_pc1_graph}. Additional details for this particular protein are provided in SI Fig.~\ref{fig_SI:ntl9_bruschweiller}.  This analysis reveals a trend that differs markedly from ubiquitin. In ubiquitin, \(\mathrm{PC}_1\) is the most delocalized mode, with contributions distributed across the protein. By contrast, in NTL9, \(\mathrm{PC}_1\) is comparatively localized, with a collectivity index of approximately \(0.4\). The collectivity indices of \(\mathrm{PC}_{2,3,4,5}\) are \(0.52\), \(0.38\), \(0.29\), and \(0.65\), respectively. Structurally, \(\mathrm{PC}_1\) is primarily localized around the \(\beta\)-turn region of the protein, and \(\mathrm{PC}_{2}\), \(\mathrm{PC}_{3}\), and \(\mathrm{PC}_{4}\) also involve motions broadly concentrated in this region, with additional secondary contributions. This suggests that the strong information transfer from \(\mathrm{PC}_1\) to these modes reflects a local hierarchy of motions, in which a slow fluctuation centered around the \(\beta\)-turn organizes faster fluctuations in the same structural region.

A particularly interesting case is \(\mathrm{PC}_5\), which is substantially more delocalized than the other higher-order modes considered here. Despite this difference in spatial character, the \(\mathrm{PC}_1 \rightarrow \mathrm{PC}_5\) link remains strongly asymmetric. This observation indicates that directional information transfer between PCs is not necessarily determined by localization or delocalization. In the case of NTL9, a localized slow fluctuation can carry predictive information about a more delocalized mode involving other parts of the protein. This result highlights the nontrivial character of information transfer in the PC representation and suggests that the present framework may provide a useful route for identifying dynamical couplings between motions occurring in distinct regions of a protein, with potential relevance for future studies of allosteric communication.

\section{Discussion} \label{sec:discussion}

In this work, we have investigated the presence of putative causal links between  principal components extracted from long equilibrium protein trajectories. Principal component analysis is most often used as a dimensionality-reduction tool, where the leading modes are interpreted in terms of variance, collectivity, and structural deformation. Here, we have taken a complementary perspective and asked whether the present value of one PC improves the prediction of the future evolution of another PC. By using the Imbalance Gain approach\cite{deltatto2024robust}, we obtain a time-lagged and directional measure of information transfer between collective modes. Importantly, the directional information transfer identified by our approach cannot be identified by the TICA analysis, since the time-delayed covariance matrix between PCs is symmetric if the dynamics satisfies detailed balance, as is the case for molecular dynamics. 

The two systems studied here display qualitatively distinct patterns of directional coupling. In ubiquitin, the leading principal component, which is also the slowest one, is a highly delocalized, protein-wide mode and exhibits the strongest information transfer toward \(\mathrm{PC}_2\) and \(\mathrm{PC}_3\). The influence of \(\mathrm{PC}_1\) becomes progressively weaker and more symmetric for higher-order PCs, indicating that in this case the dominant collective fluctuation primarily influences  a few leading modes. This behavior is consistent with a picture in which a broad native-state fluctuation provides a slowly varying dynamical context for more localized motions, while lower-variance modes become increasingly weakly coupled to the dominant PC.

In NTL9, \(\mathrm{PC}_1\) is clearly the slowest mode and displays a pronounced directional influence on all the other leading PCs considered. Unlike ubiquitin, however, this dominant mode is not the most delocalized one. Instead, \(\mathrm{PC}_1\) is  localized around the \(\beta\)-turn region, while some of the modes receiving information from it, in particular \(\mathrm{PC}_5\), are more spatially extended. This result shows that directional information transfer between PCs is not simply determined by the collectivity or variance of the corresponding modes. A localized slow fluctuation can carry predictive information about faster and more delocalized motions, suggesting that the causal organization of PC space can reveal dynamical relationships that are not apparent from the PCA spectrum or collectivity indices alone.

It is important to remark that PCA and the IG  between PCs probe two different aspects. PCA identifies directions of large-amplitude structural variance, whereas IG asks whether fluctuations along one such direction improve the prediction of another at later times. The first criterion is geometric and instantaneous, while the second is dynamical and time-lagged. Therefore, the most informative causal driver need not be the PC with the largest variance or the most spatially delocalized eigenvector. In ubiquitin, these properties approximately coincide: the most collective mode is also the dominant source of directional information transfer. In NTL9, they separate: the slowest and most predictive mode is relatively localized, yet it influences both localized and more distributed fluctuations. This distinction highlights the value of studying causal structure in PC space rather than relying exclusively on variance, relaxation times, or spatial collectivity. In the systems analyzed so far, we observe several cases in which the IGT values are broadly symmetric between pairs of PCs (e.g., between $\mathrm{PC}_2 \Leftrightarrow \mathrm{PC}_{3,4,5}$ in NTL9, see SI Fig.~\ref{fig_SI:ntl9_pc2_graph}). However, we have not yet identified a case in which the dominant causal direction is inverted relative to the ordering of explained variance; that is, we do not observe lower-variance PCs acting as stronger causal drivers of higher-variance PCs.


The present results suggest several natural extensions. One direction is to move from pairwise causal graphs toward multivariate  formulations in which the influence of a PC is assessed while accounting for the simultaneous effects of other modes. The approach introduced in Ref. \citenum{allione2025linear} might allow us to perform this analysis, with the caveat that the statistical uncertainty of the inferred collective links should be carefully controlled.  Performing this analysis would help distinguish whether the observed links correspond to direct dynamical couplings or are mediated by intermediate PCs. A second direction is to connect PC-level causality to residue-level or structural mechanisms, for example by combining IG with mode localization, contact maps, or residue-wise dynamical descriptors. This would be particularly useful for systems such as NTL9, where a localized slow fluctuation appears to influence more extended protein motions. Finally, the framework could be applied to proteins with known allosteric regulation, ligand binding, or conformational switching, where directional couplings between collective modes may provide a compact description of how local perturbations propagate through the protein.\cite{rivalta2012allosteric,vogt2012conformational,vanwart2012exploring,guo2016protein,hertig2016revealing,bowerman2016detecting,rivalta2016allosteric,wang2020mapping,hacisuleyman2017entropy,wodak2019allostery,hardie2023deconstructing,bernetti2024probing}




\section*{Supplementary Material}
The Supplementary Material includes a discussion of the different MD trajectories that are analyzed, a description of the Brüschweiler collectivity index, results for NTL9 from another trajectory fragment that show the consistency of the results, and supplementary figures that show the PCA eigenvalue spectra, the cosine similarity of PCs between two halves of the analyzed trajectories to show consistency of the PCs across both halves, the Brüschweiler collectivity index plots, the IG curves from \(\mathrm{PC}_1 \Leftrightarrow \mathrm{PC}_{6,7,8,9}\), and the IG curves and corresponding IGT graphs from \(\mathrm{PC}_2 \Leftrightarrow \mathrm{PC}_{3,4,5}\) for both systems.


\begin{acknowledgments}
DB and AH thank the European Commission for funding from the ERC Grant HyBOP 101043272. 
This work was partially funded by NextGenerationEU through the Italian National Centre for HPC, Big Data, and Quantum Computing (Grant No. CN00000013 received by A.L.). The authors thank D. E. Shaw Research for sharing their ubiquitin trajectory from Ref.~\citenum{DEShaw2016} and their NTL9 trajectory from Ref.~\citenum{DEShaw2011}. The authors also acknowledge the use of AI tools to improve the writing and readability of the manuscript.
\end{acknowledgments}


\section*{Data Availability Statement}

The data that support the findings of this study are openly available at \url{https://github.com/debarshibanerjee/pca_causality_proteins}. The trajectories were provided by D. E. Shaw Research.


\bibliography{refs}



\usetikzlibrary{arrows.meta, positioning, fit, backgrounds, calc, shapes, shapes.geometric}

\DeclareSIUnit\angstrom{\text {Å}}
\pdfstringdefDisableCommands{\let\ce\relax}



\newcommand{\beginsupplement}{%
        \setcounter{table}{0}
        \renewcommand{\thetable}{S\arabic{table}}%
        \setcounter{figure}{0}
        \renewcommand{\thefigure}{S\arabic{figure}}%
        \setcounter{equation}{0}
        \renewcommand{\theequation}{S\arabic{equation}}%
        \setcounter{section}{0}
        \renewcommand{\thesection}{S\arabic{section}}%
        \setcounter{subsection}{0}
        \renewcommand{\thesubsection}{S\arabic{subsection}}%
        \newcounter{SItab}
        \renewcommand{\theSItab}{S\arabic{SItab}}%
        \newcounter{SIfig}
        \renewcommand{\theSIfig}{S\arabic{SIfig}}%
}









\clearpage
\beginsupplement

\section*{Supplementary Information}

\subsection{Analysis of trajectories} \label{sec_si:md_traj}
For both systems, the use of long molecular dynamics trajectories is essential to ensure that the principal components are statistically converged and representative of the underlying folded-state dynamics. Convergence was assessed by dividing each selected trajectory into two halves, independently computing the principal components for each half, and evaluating their consistency through the cosine similarity between the corresponding eigenvectors. Refer to SI Figs.~\ref{fig_SI:ubq_pc_cosine_sim} and \ref{fig_SI:ntl9_pc_cosine_sim} for the results of this calculation for Ubiquitin and NTL9 respectively.
The observation that the same dominant PCs are recovered from both halves provides a stringent check that the PCA is not affected by insufficient sampling.

\subsubsection*{Ubiquitin}

The ubiquitin trajectory analyzed in this work corresponds to the \(1~\mathrm{ms}\) simulation reported in Ref.~\citenum{DEShaw2016}. In that study, the RMSD was computed over residues \(2\)--\(71\), thereby excluding the flexible C-terminal tail. We therefore restricted our analysis to the same residue range. Moreover, the trajectory exhibits a conformational transition at approximately \(0.65~\mathrm{ms}\), as reflected in the RMSD profile shown in Fig.~1 of Ref.~\citenum{DEShaw2016}. To avoid mixing distinct conformational regimes, we limited the present analysis to the first \(650~\mu\mathrm{s}\) of the simulation.

\subsubsection*{NTL9}

The NTL9 protein was simulated for a total of \(2.9~\mathrm{ms}\) in Ref.~\citenum{DEShaw2011}, distributed across four independent simulations. Full details of these simulations are provided in the Supporting Information of Ref.~\citenum{DEShaw2011}. Among these trajectories, the first two have lengths of \(1.112~\mathrm{ms}\) and \(1.073~\mathrm{ms}\), respectively, and each contains two long folded-state segments exceeding \(200~\mu\mathrm{s}\). Since our objective was to analyze principal components representative of the folded-state dynamics, we selected two such folded trajectory segments for the present analysis. The results reported in the main text were obtained from the segment spanning \(830\)--\(1060~\mu\mathrm{s}\) of the \(1.073~\mathrm{ms}\) trajectory, whereas the consistency check presented below was performed using the segment spanning \(290\)--\(580~\mu\mathrm{s}\) of the \(1.112~\mathrm{ms}\) trajectory.

\subsection{Brüschweiler collectivity index} \label{sec_si:bruschweiller}

To quantify the spatial extent of each principal component (PC), we computed the
Brüschweiler collectivity index.\cite{Bruschweiler1995,Bruschweiler2000} 
In the present analysis, PCA was performed using only the C\(_\alpha\) atoms. Therefore, each C\(_\alpha\) atom was associated with a single residue, and the collectivity index can be interpreted as a residue-level measure of how broadly a given PC mode is distributed over the protein.

For a protein containing \(N\) C\(_\alpha\) atoms, the eigenvector associated with PC\(_k\)
is written as
\[
\mathbf{q}_k =
\left(
q_{k,1x},q_{k,1y},q_{k,1z},
q_{k,2x},q_{k,2y},q_{k,2z},
\ldots,
q_{k,Nx},q_{k,Ny},q_{k,Nz}
\right),
\]
where \((q_{k,ix},q_{k,iy},q_{k,iz})\) denotes the three-dimensional displacement
component of residue \(i\) along mode \(k\). The squared amplitude associated with
residue \(i\) was computed as
\[
a_{k,i}
=
q_{k,ix}^{2}
+
q_{k,iy}^{2}
+
q_{k,iz}^{2}.
\]
The normalized residue-wise contribution to mode \(k\) was then defined as
\[
p_{k,i}
=
\frac{a_{k,i}}{\sum_{j=1}^{N} a_{k,j}},
\qquad
\sum_{i=1}^{N} p_{k,i}=1.
\]
Thus, \(p_{k,i}\) represents the fraction of the total squared amplitude of PC\(_k\)
carried by residue \(i\).

The Brüschweiler collectivity index of mode \(k\) was calculated as
\[
\kappa_k
=
\frac{1}{N}
\exp
\left[
-\sum_{i=1}^{N}
p_{k,i}\log p_{k,i}
\right].
\]

The index ranges approximately from \(1/N\) to \(1\). A value close to \(1/N\)
indicates a highly localized mode, in which the displacement is concentrated on a
single residue or a small number of residues. Conversely, a value close to \(1\)
indicates a collective mode, in which the displacement amplitude is broadly and
approximately uniformly distributed over the protein. In this work, larger values of
\(\kappa_k\) were therefore interpreted as more delocalized, collective PC modes,
whereas smaller values indicated more localized modes.

\subsection{NTL9 results for another folded trajectory fragment}
\label{sec_si:ntl9_extra_results}

To assess the robustness of the causal relationships inferred for NTL9, we repeated the analysis on an independent folded-state trajectory fragment. Specifically, we considered the segment spanning \(290\)--\(580~\mu\mathrm{s}\) of the \(1.112~\mathrm{ms}\) trajectory, during which the protein remains in the folded state. This analysis provides an additional consistency check for the results reported in the main text, which were obtained from a distinct folded segment of the NTL9 trajectory.

The autocorrelation functions of the first five principal components are shown in Fig.~\ref{fig_SI:ntl9-0_acf}. They display the same qualitative behavior observed for the trajectory fragment analyzed in the main text: \(\mathrm{PC}_1\) is the slowest relaxing mode, while the remaining leading PCs decay on shorter time scales. This indicates that the separation between the dominant slow mode and the higher-order PCs is not specific to a single folded-state segment.

We next computed the pairwise IG between \(\mathrm{PC}_1\) and \(\mathrm{PC}_{2,3,4,5}\), shown in Fig.~\ref{fig_SI:ntl9-0_ig_curves}. The resulting IG curves again reveal a pronounced directional asymmetry, with the dominant information transfer oriented from \(\mathrm{PC}_1\) toward the other leading PCs. This behavior is summarized by the causal graph in Fig.~\ref{fig_SI:ntl9-0_pc1_graph}, constructed from the normalized integrated information transfer, IGT. The graph confirms that \(\mathrm{PC}_1\) acts as the dominant source of directional information transfer in this folded trajectory fragment, consistent with the conclusions drawn from the analysis in the main text.

Finally, Fig.~\ref{fig_SI:ntl9-0_pc2_ig_curves} reports the corresponding pairwise IG estimates between \(\mathrm{PC}_2\) and \(\mathrm{PC}_{3,4,5}\). In contrast to the strongly asymmetric links involving \(\mathrm{PC}_1\), these couplings are broadly reciprocal. 
Overall, the analysis of this independent folded-state fragment supports the robustness of the main conclusion: within the folded-state dynamics of NTL9, \(\mathrm{PC}_1\) behaves as a slow mode that directionally conditions several faster principal components, whereas the causal links among the lower-variance PCs are weaker and more symmetric.

Further characterization of the leading PCs using the Br\"uschweiler collectivity index, reported in Fig.~\ref{fig_SI:ntl9-0_bruschweiller}, shows that, as in the trajectory fragment discussed in the main text, the dominant modes remain relatively localized and are mainly concentrated around the \(\beta\)-turn region. This indicates that the spatial character of the leading PCs is conserved across the two folded segments. The main difference is that, in the present fragment, the first substantially delocalized mode that is strongly driven by \(\mathrm{PC}_1\) is \(\mathrm{PC}_6\), rather than \(\mathrm{PC}_5\) as observed in the analysis in the main-text.

\subsection{Supplementary Figures}

\begin{figure}[htbp]
    \centering

    \includegraphics[width=0.8\linewidth]{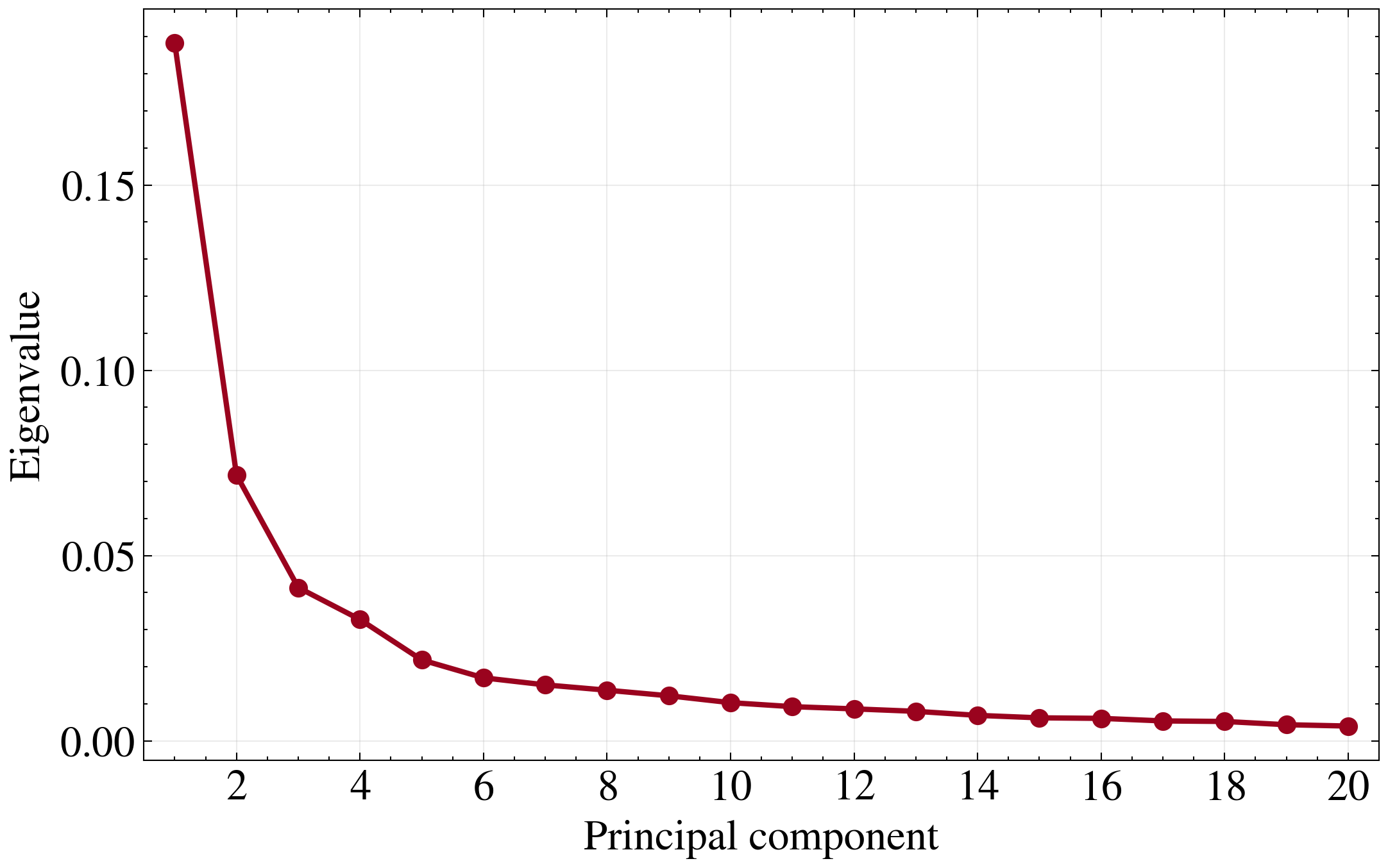}

    \vspace{0.5cm}

    \includegraphics[width=0.8\linewidth]{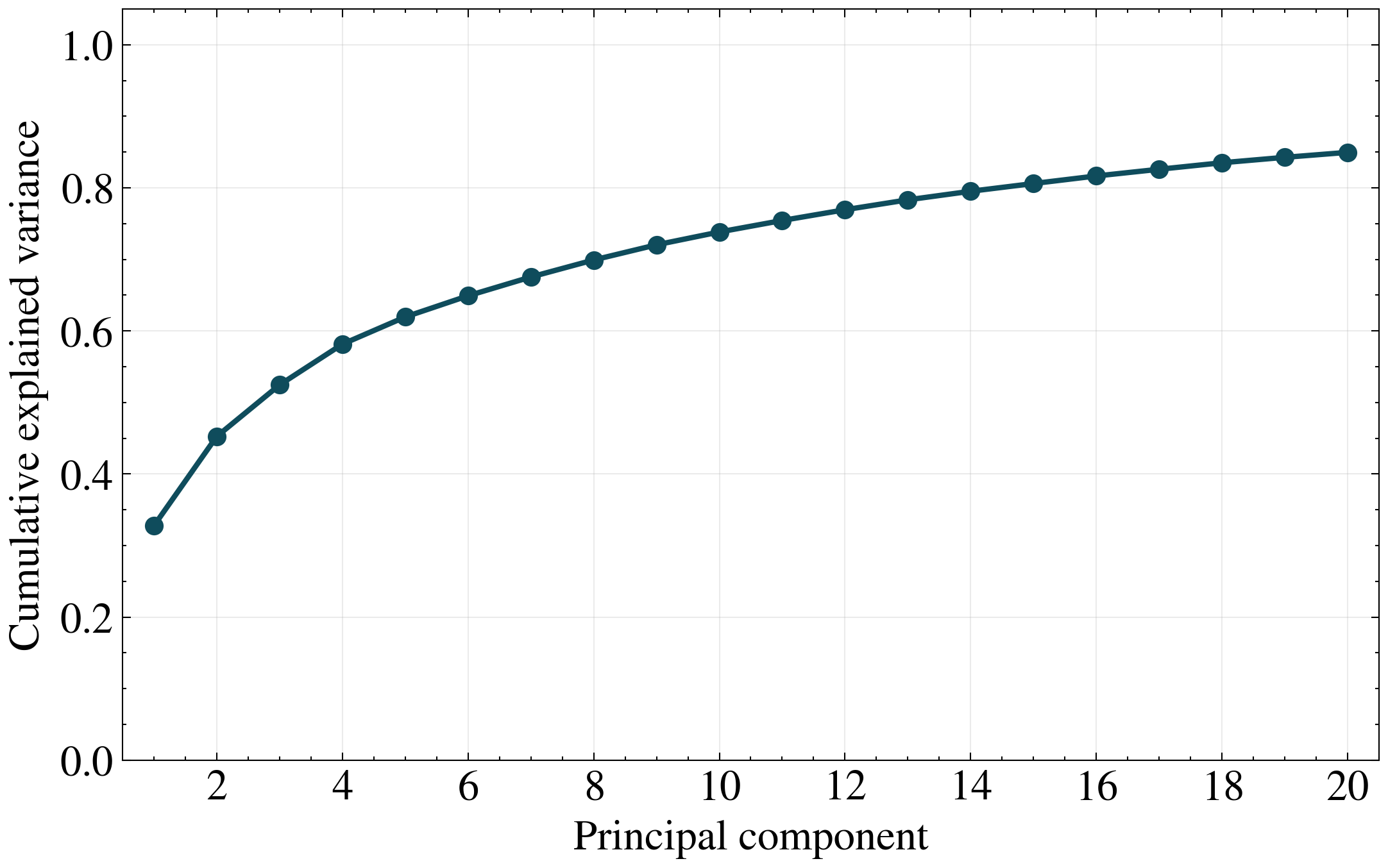}

    \caption{PCA eigenvalue spectrum and explained variance for the Ubiquitin trajectory.}
    \label{fig_SI:ubq_pca_spectrum}
\end{figure}

\begin{figure}[htbp]
  \centering
  \includegraphics[width=0.95\linewidth]{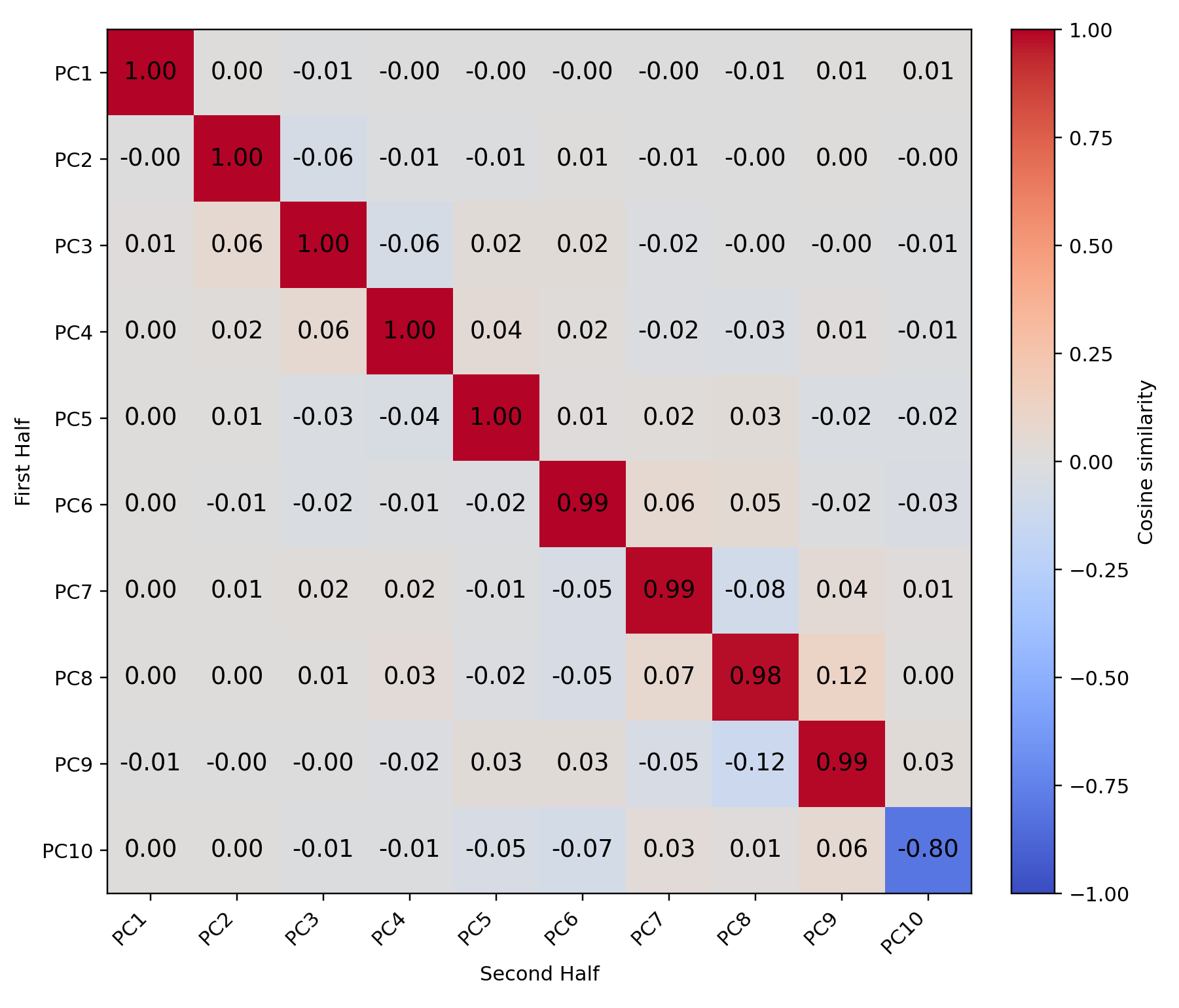}
\caption{Ubiquitin: Cosine similarity between the first ten PCs computed independently from the two halves of the analyzed ubiquitin trajectory. The near-diagonal structure indicates that the dominant PCs are consistently recovered in both halves, confirming that the PCA is converged and that the leading modes retain the same character across the trajectory.
}
  \label{fig_SI:ubq_pc_cosine_sim}
\end{figure}

\begin{figure}[htbp]
    \centering

    \includegraphics[width=0.8\linewidth]{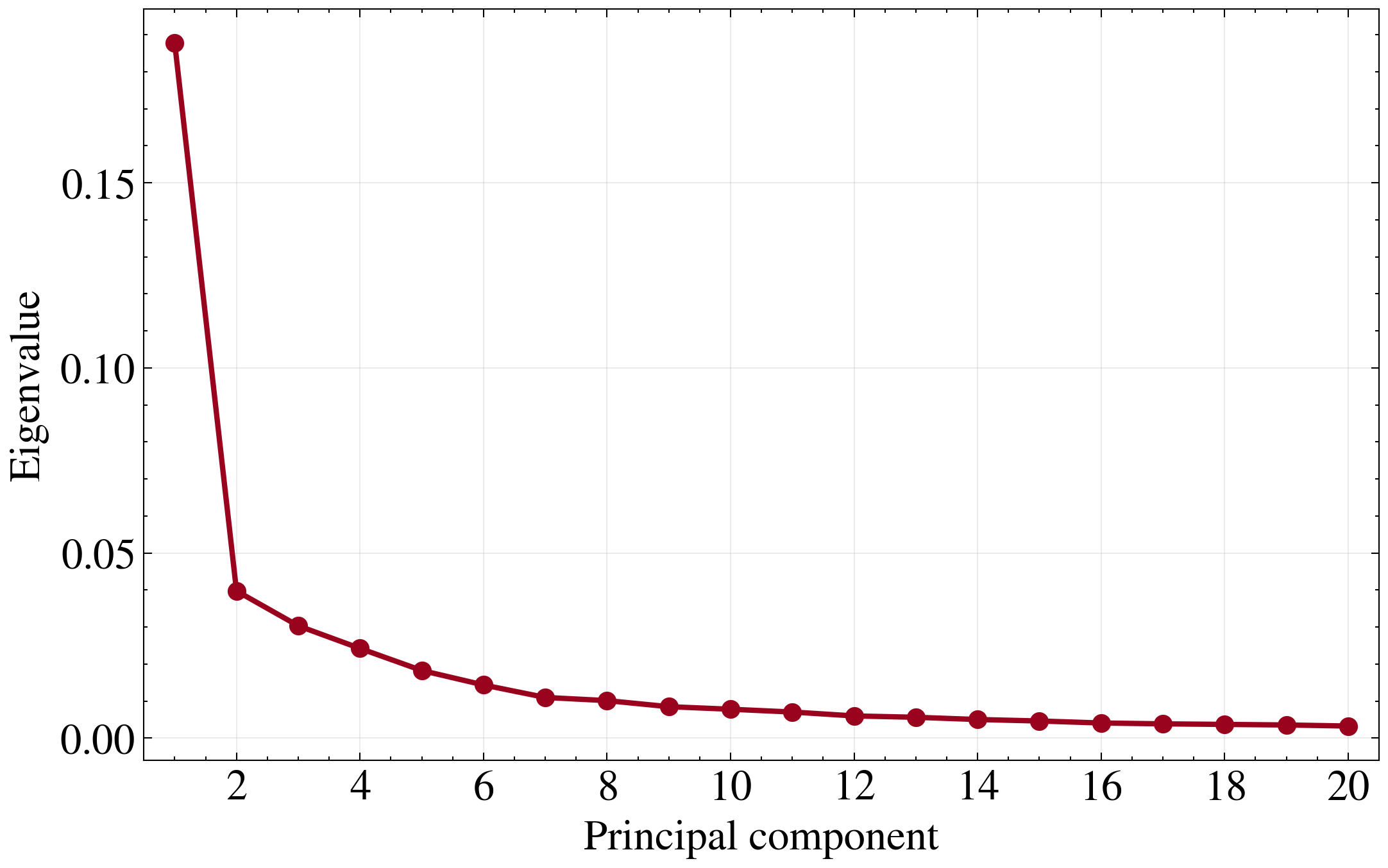}

    \vspace{0.5cm}

    \includegraphics[width=0.8\linewidth]{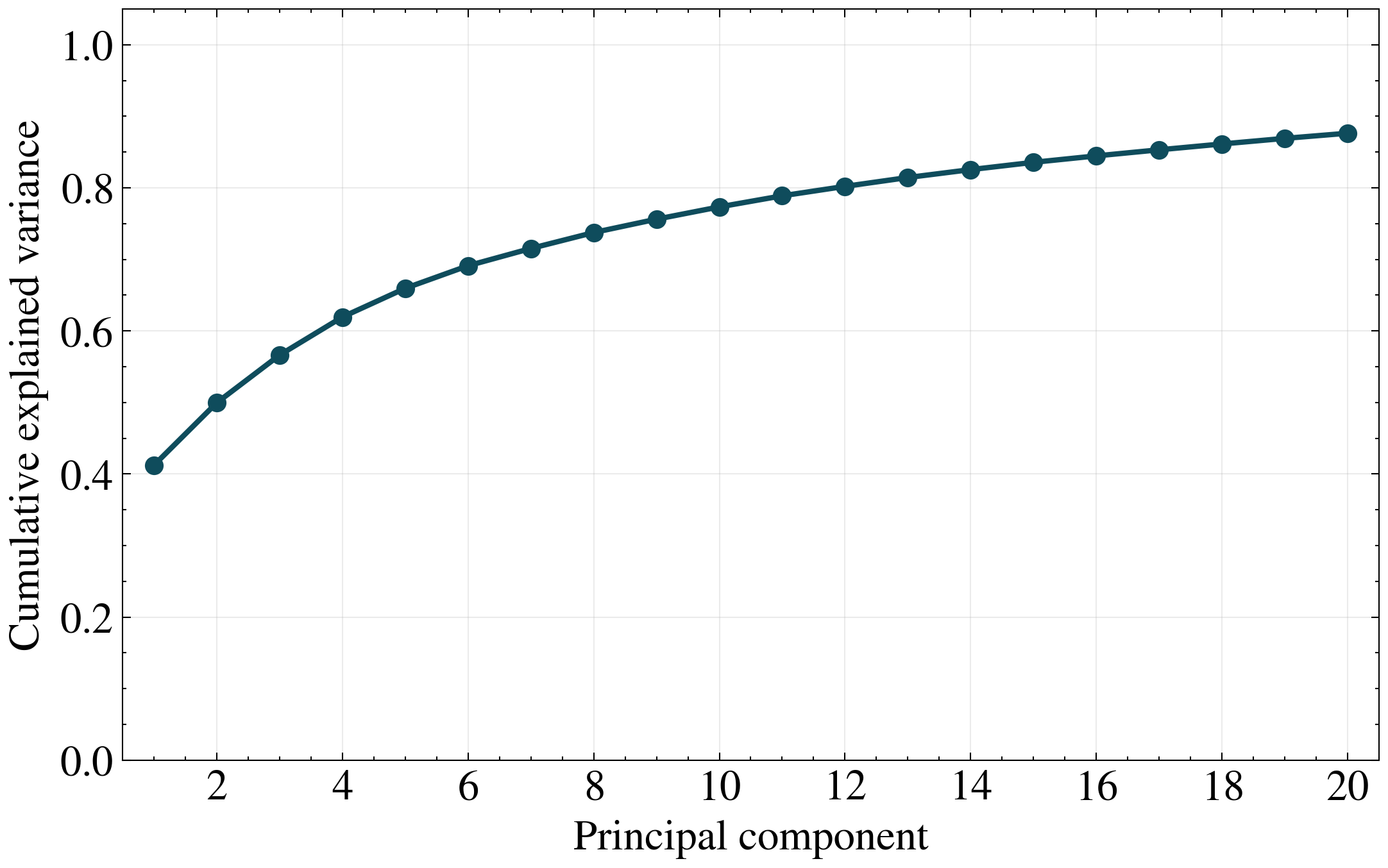}

    \caption{PCA eigenvalue spectrum and explained variance for the NTL9 trajectory.}
    \label{fig_SI:ntl9_pca_spectrum}
\end{figure}

\begin{figure}[htbp]
  \centering
  \includegraphics[width=0.95\linewidth]{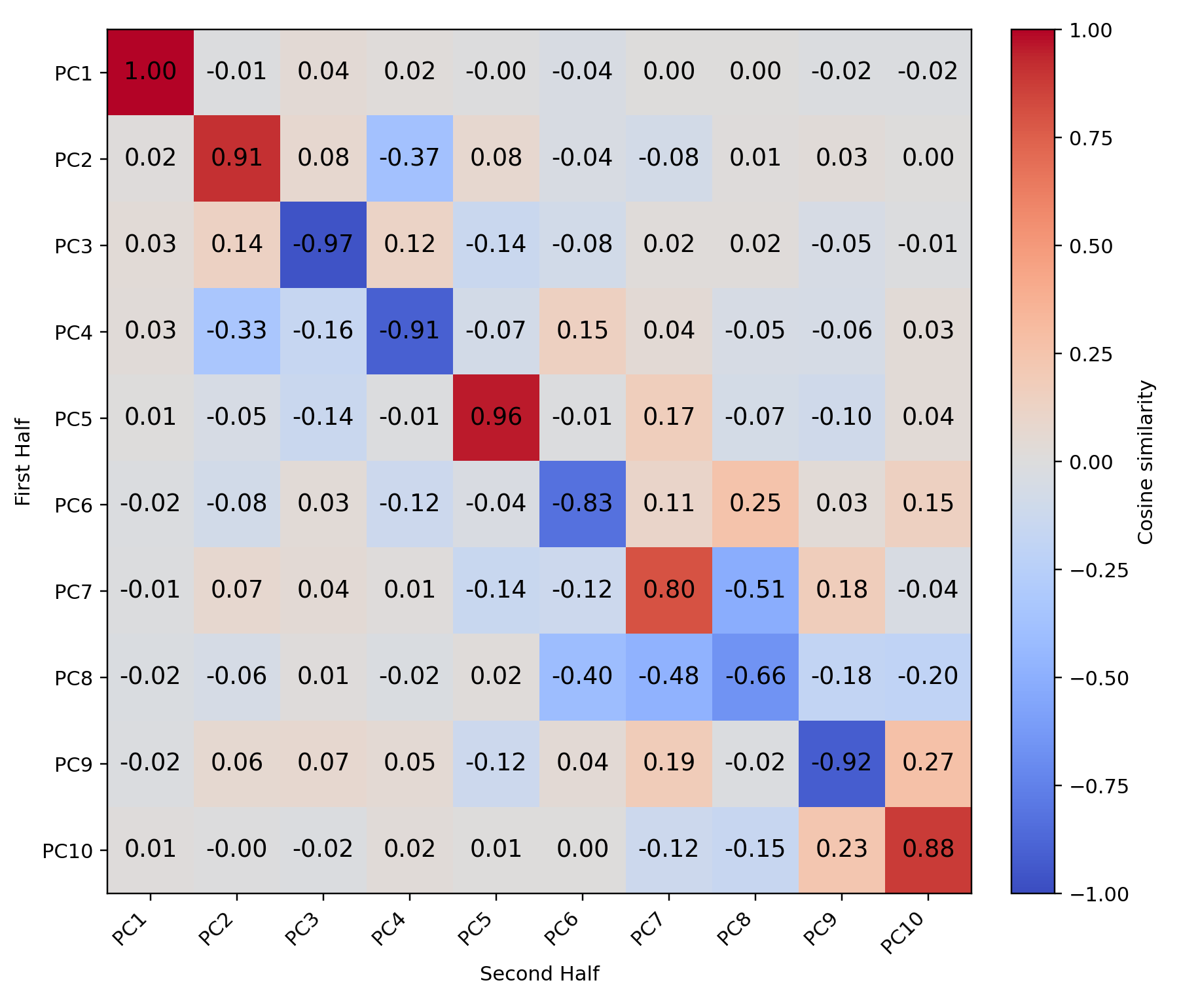}
\caption{NTL9: Cosine similarity between the first ten PCs computed independently from the two halves of the analyzed NTL9 trajectory. The near-diagonal structure, especially for the top 5 PCs, indicates that the dominant PCs are consistently recovered in both halves, confirming that the PCA is converged and that the leading modes retain the same character across the trajectory.
}
  \label{fig_SI:ntl9_pc_cosine_sim}
\end{figure}

\begin{figure}[htbp]
  \centering
  \includegraphics[width=1.0\linewidth]{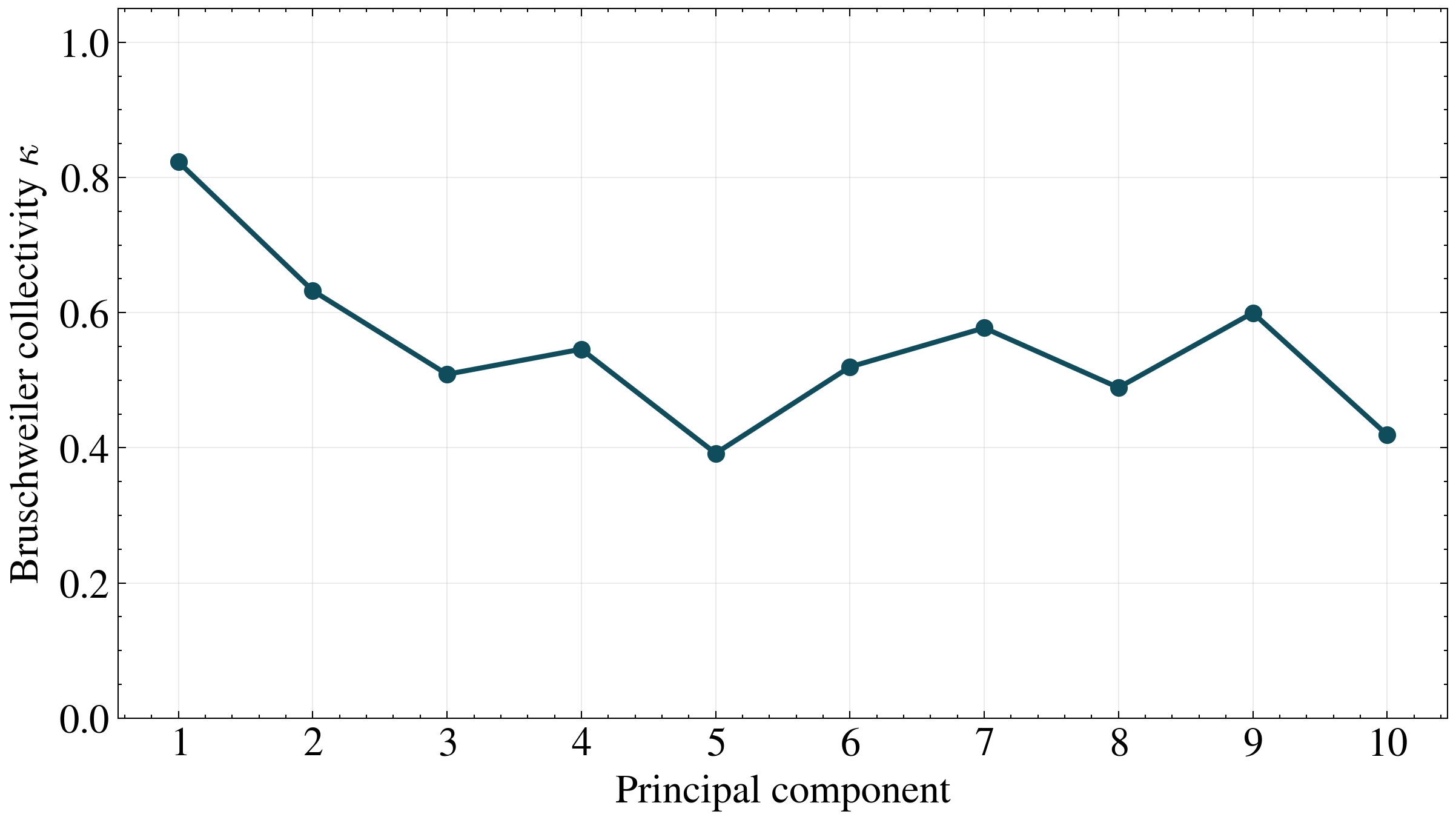}
\caption{Ubiquitin: Br\"uschweiler collectivity index for the leading PCs. The collectivity index quantifies the spatial extent of each mode.
}
  \label{fig_SI:ubq_bruschweiller}
\end{figure}

\begin{figure}[htbp]
  \centering
  \includegraphics[width=1.0\linewidth]{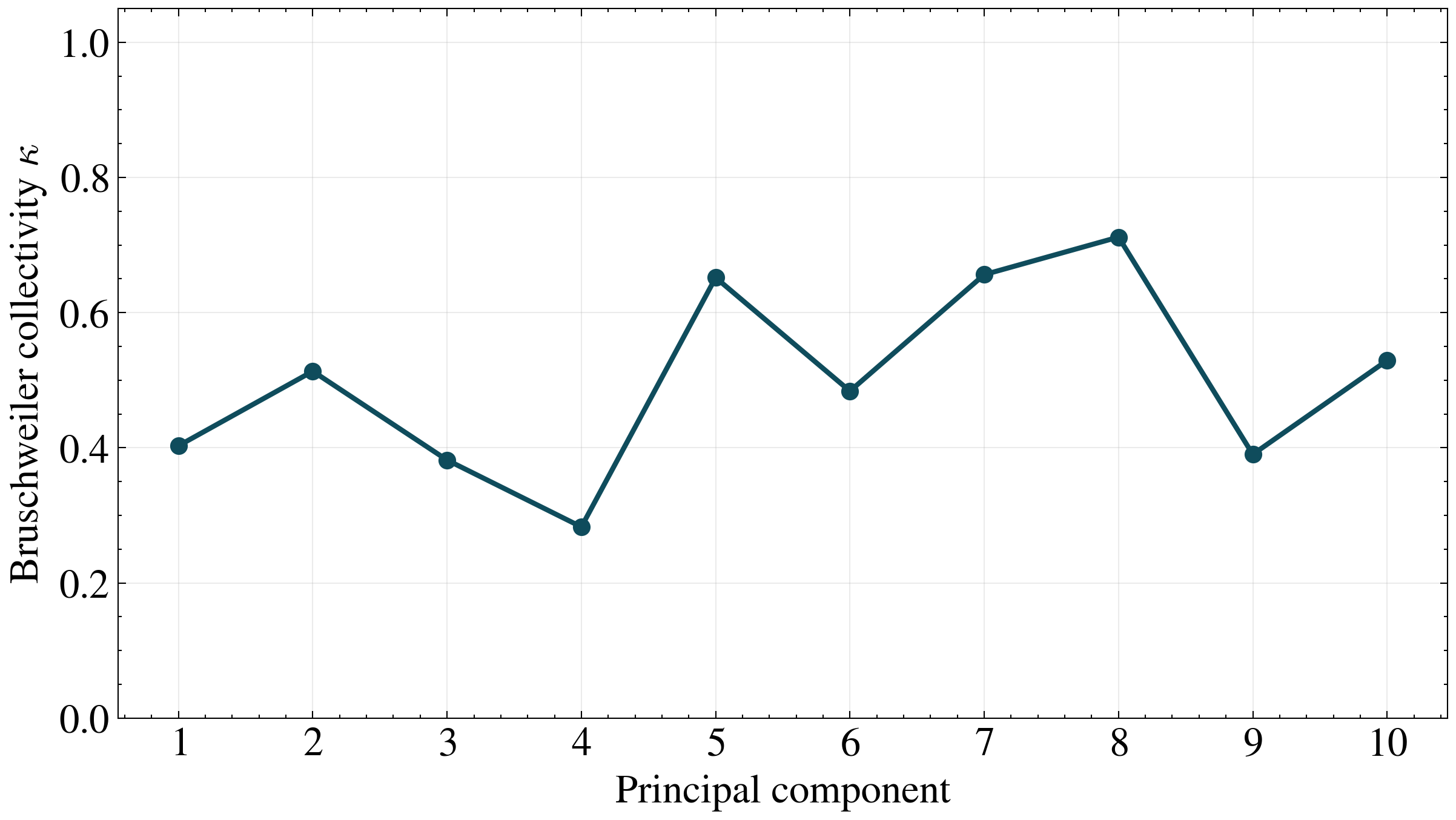}
\caption{NTL9: Br\"uschweiler collectivity index for the leading PCs. The collectivity index quantifies the spatial extent of each mode.
}
  \label{fig_SI:ntl9_bruschweiller}
\end{figure}

\begin{figure}[htbp]
  \centering
  \includegraphics[width=1.0\linewidth]{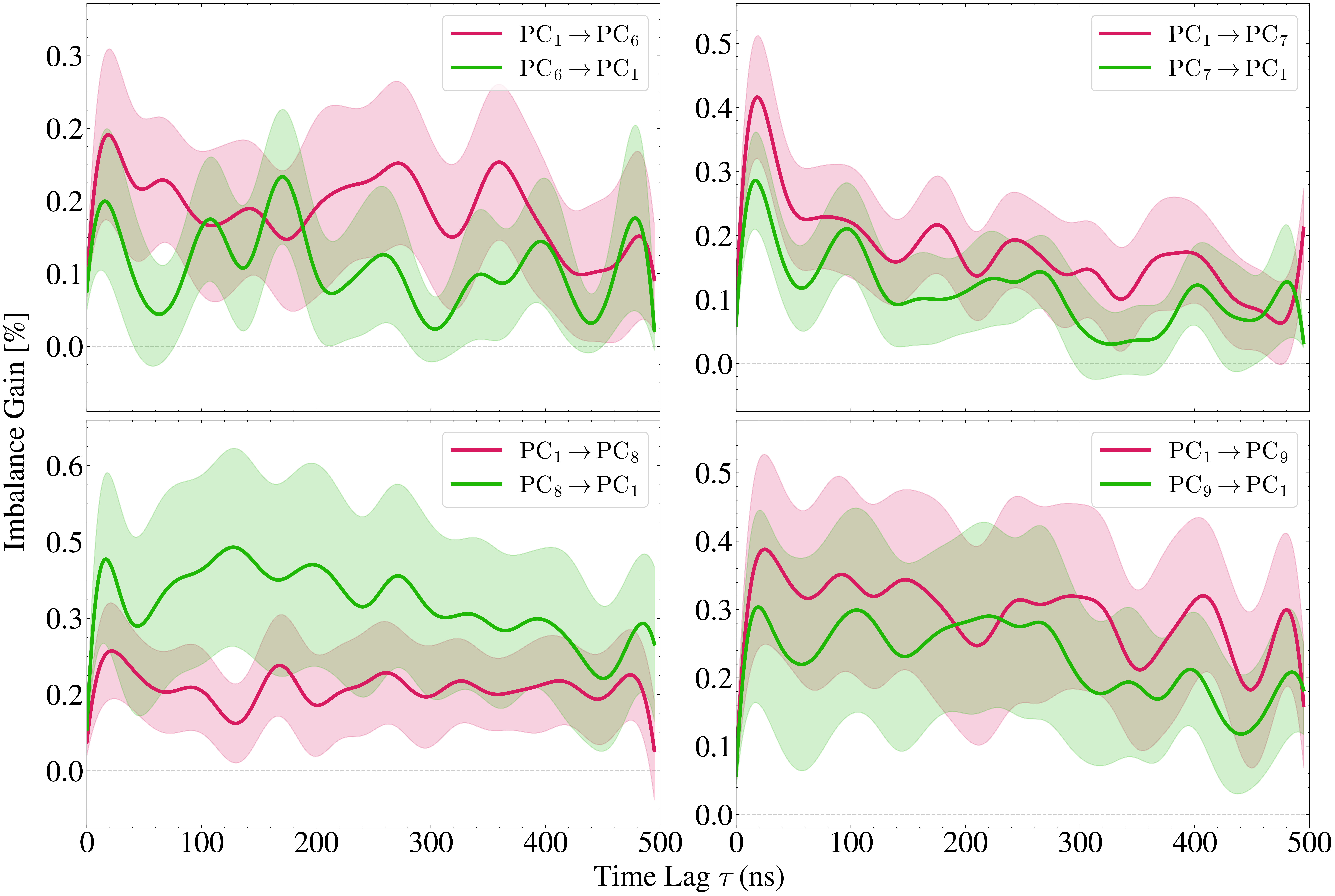}
\caption{Ubiquitin: Pairwise IG curves between \(\mathrm{PC}_1 \Leftrightarrow \mathrm{PC}_{6,7,8,9}\). The IG values are on the Y-axis, and time lag ($\tau$) in nanoseconds is on the X-axis. Pink lines denote the direction $\mathrm{PC}_1 \rightarrow \mathrm{PC}_{6,7,8,9}$, whereas green lines denote the reverse direction. Shaded areas denote error bars computed over 20 independent realizations.
}
  \label{fig_SI:ubq_pc1_rest_ig}
\end{figure}

\begin{figure}[htbp]
  \centering
  \includegraphics[width=1.0\linewidth]{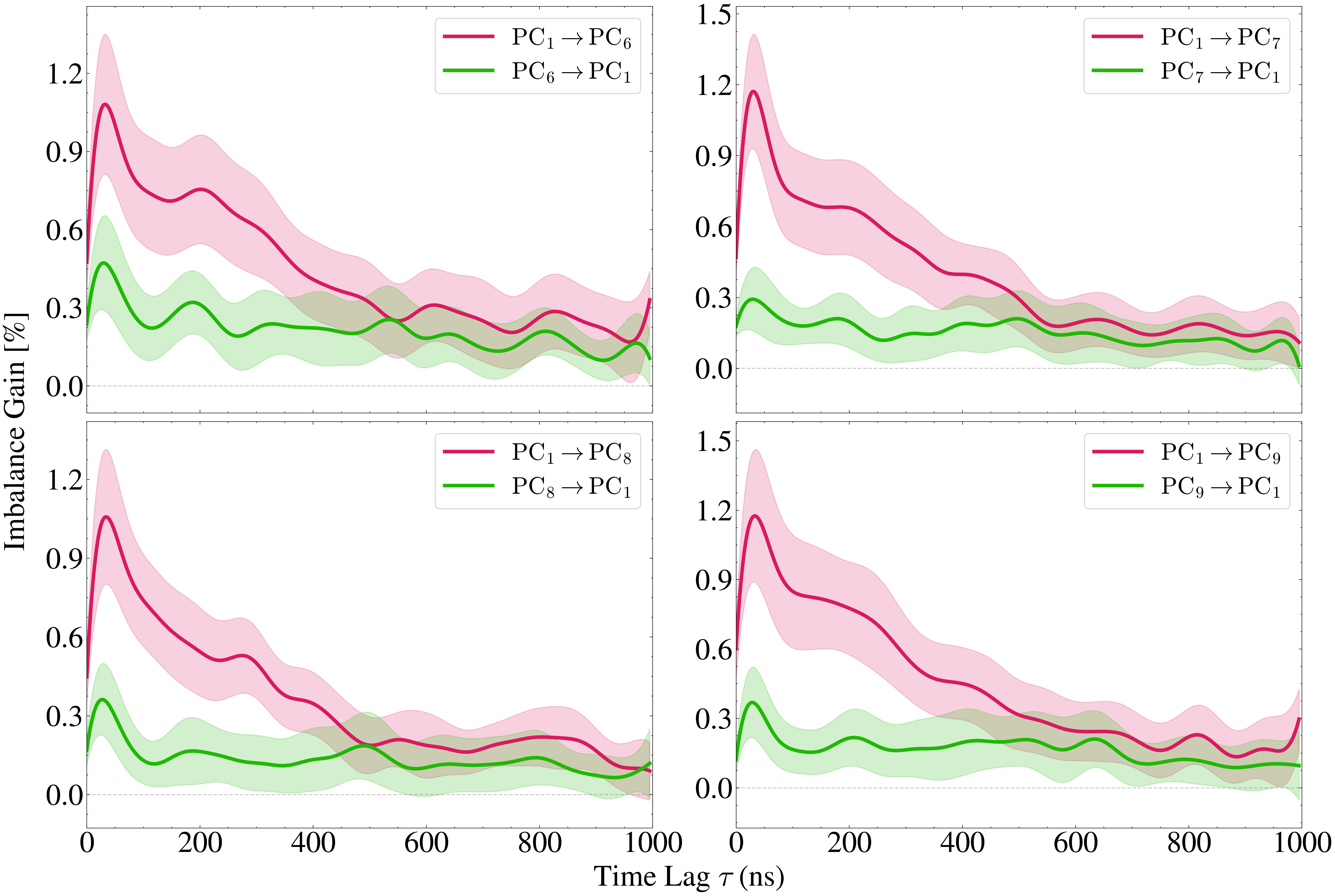}
\caption{NTL9: Pairwise IG curves between \(\mathrm{PC}_1 \Leftrightarrow \mathrm{PC}_{6,7,8,9}\). The IG values are on the Y-axis, and time lag ($\tau$) in nanoseconds is on the X-axis. Pink lines denote the direction $\mathrm{PC}_1 \rightarrow \mathrm{PC}_{6,7,8,9}$, whereas green lines denote the reverse direction. Shaded areas denote error bars computed over 20 independent realizations.
}
  \label{fig_SI:ntl9_pc1_rest_ig}
\end{figure}

\begin{figure}[htbp]
\centering

\includegraphics[width=1.0\linewidth]{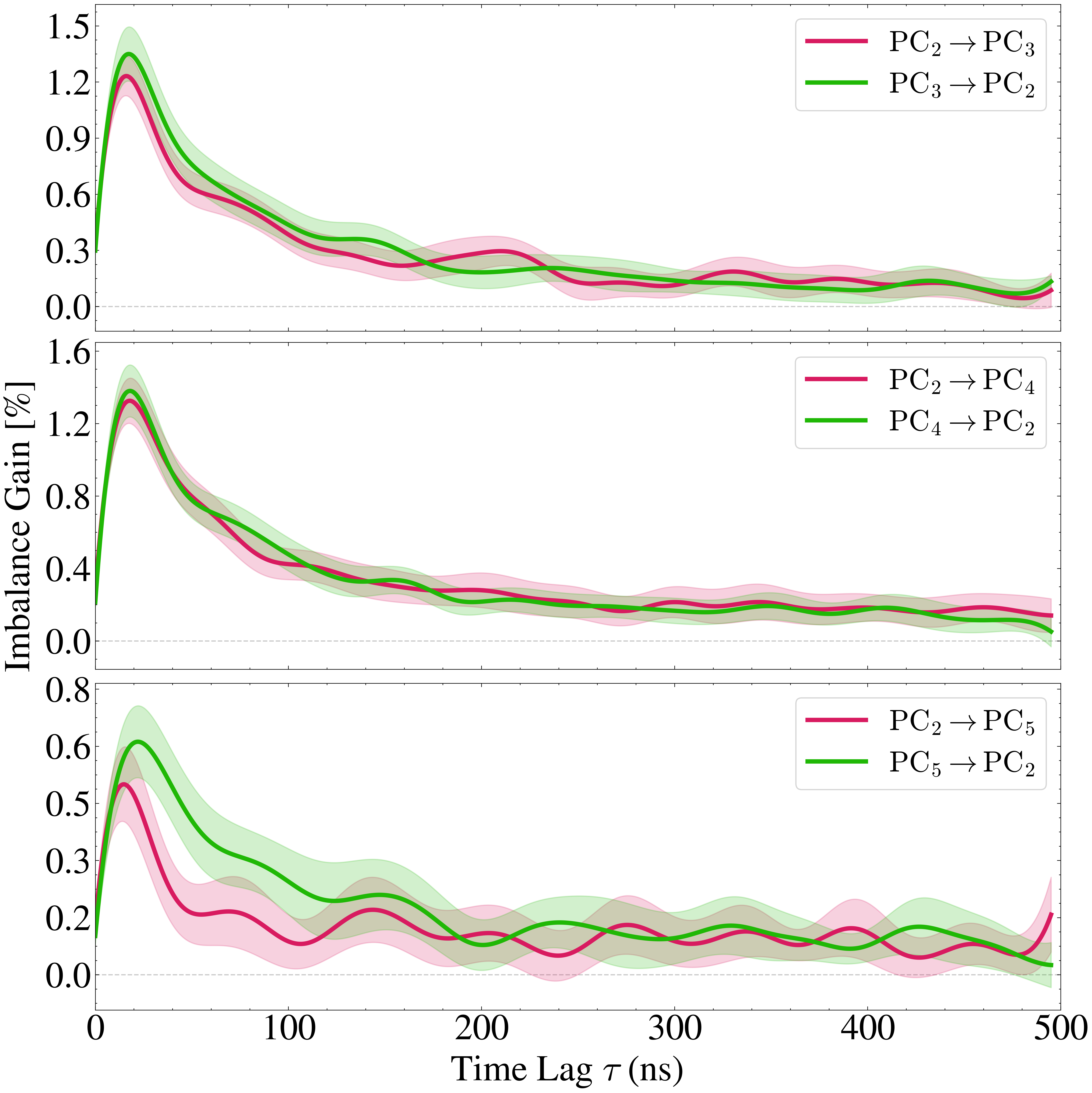}

\vspace{0.5cm}

\resizebox{1.0\linewidth}{!}{
\begin{tikzpicture}
[
    PC/.style={circle, draw=black!70, fill=black!5, minimum size=1.8cm, font=\Large\sffamily, thick},
    PCcenter/.style={circle, draw=red!70!black, fill=red!20, minimum size=1.8cm, font=\Large\sffamily, thick}
]


    \node[PCcenter] (PC2) at (0,0) {PC$_2$};

    \def\R{7} 

    \node[PC] (PC3) at (210:\R) {PC$_3$};
    \node[PC] (PC4) at (270:\R) {PC$_4$};
    \node[PC] (PC5) at (330:\R) {PC$_5$};

    \begin{scope}[on background layer]
        \foreach \i in {3,4,5} {
            \draw[loosely dotted, very thick, purple!50!gray] (PC2) -- (PC\i);
        }
    \end{scope}

    \newcommand{\proparrow}[4]{
        \path (#1) -- (#2) coordinate[pos=0] (bStart) coordinate[pos=1] (bEnd);
        
        \pgfmathsetmacro{\vfrac}{#3 * 0.5}
        
        \draw[-{Latex[length=4.5mm, width=3.5mm]}, line width=2.5pt, draw=black!90]
            (bStart) -- ($ (bStart) ! \vfrac ! (bEnd) $)
            node[midway, #4, font=\normalsize\sffamily, inner sep=5pt] {#3};
    }

    
    
    \proparrow{PC2}{PC5}{0.20}{sloped, above}
    \proparrow{PC5}{PC2}{0.20}{sloped, below}

    \proparrow{PC2}{PC4}{0.63}{sloped, above}
    \proparrow{PC4}{PC2}{0.38}{sloped, below}

    \proparrow{PC2}{PC3}{0.87}{sloped, above}
    \proparrow{PC3}{PC2}{0.60}{sloped, below}

\end{tikzpicture}
} 

\caption{Ubiquitin: The top panel shows the pairwise IG curves between \(\mathrm{PC}_2 \Leftrightarrow \mathrm{PC}_{3,4,5}\). The IG values are on the Y-axis, and time lag ($\tau$) in nanoseconds is on the X-axis. Pink lines denote the direction $\mathrm{PC}_2 \rightarrow \mathrm{PC}_{3,4,5}$, whereas green lines denote the reverse direction. Shaded areas denote error bars computed over 20 independent realizations. The bottom panel shows the putative causal links summarizing the normalized IGT between \(\mathrm{PC}_2 \Leftrightarrow \mathrm{PC}_{3,4,5}\). Arrow lengths are proportional to the magnitude of the IGT, and the corresponding IGT values are reported next to each arrow. \(\mathrm{PC}_2\) exhibits moderately asymmetric causal links with \(\mathrm{PC}_3\) and \(\mathrm{PC}_4\), with weaker information transfer in the reverse directions. In contrast, the link between \(\mathrm{PC}_2\) and \(\mathrm{PC}_5\) is weak and approximately bidirectional.
}

\label{fig_SI:ubq_pc2_graph}
\end{figure}

\begin{figure}[htbp]
\centering

\includegraphics[width=1.0\linewidth]{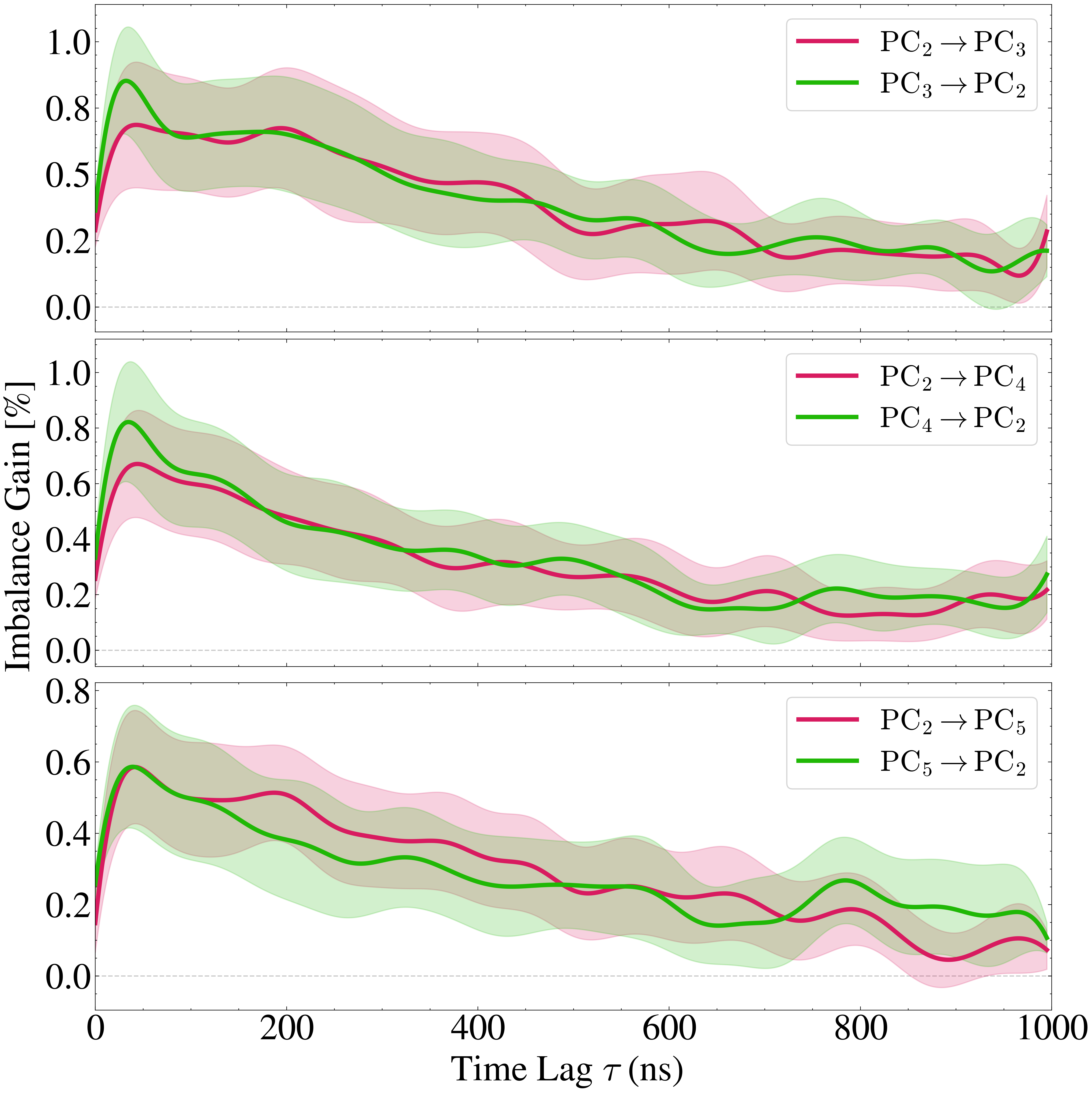}

\vspace{0.5cm}

\resizebox{1.0\linewidth}{!}{
\begin{tikzpicture}[
    PC/.style={circle, draw=black!70, fill=black!5, minimum size=1.8cm, font=\Large\sffamily, thick},
    PCcenter/.style={circle, draw=red!70!black, fill=red!20, minimum size=1.8cm, font=\Large\sffamily, thick}
]


    \node[PCcenter] (PC2) at (0,0) {PC$_2$};

    \def\R{7} 

    \node[PC] (PC3) at (210:\R) {PC$_3$};
    \node[PC] (PC4) at (270:\R) {PC$_4$};
    \node[PC] (PC5) at (330:\R) {PC$_5$};

    \begin{scope}[on background layer]
        \foreach \i in {3,4,5} {
            \draw[loosely dotted, very thick, purple!50!gray] (PC2) -- (PC\i);
        }
    \end{scope}

    \newcommand{\proparrow}[4]{
        \path (#1) -- (#2) coordinate[pos=0] (bStart) coordinate[pos=1] (bEnd);
        
        \pgfmathsetmacro{\vfrac}{#3 * 0.5}
        
        \draw[-{Latex[length=4.5mm, width=3.5mm]}, line width=2.5pt, draw=black!90]
            (bStart) -- ($ (bStart) ! \vfrac ! (bEnd) $)
            node[midway, #4, font=\normalsize\sffamily, inner sep=5pt] {#3};
    }

    
    
    \proparrow{PC2}{PC5}{0.65}{sloped, above}
    \proparrow{PC5}{PC2}{0.63}{sloped, below}

    \proparrow{PC2}{PC4}{0.70}{sloped, above}
    \proparrow{PC4}{PC2}{0.74}{sloped, below}

    \proparrow{PC2}{PC3}{0.87}{sloped, above}
    \proparrow{PC3}{PC2}{0.88}{sloped, below}

\end{tikzpicture}
} 

\caption{NTL9: The top panel shows the pairwise IG curves between \(\mathrm{PC}_2 \Leftrightarrow \mathrm{PC}_{3,4,5}\). The IG values are on the Y-axis, and time lag ($\tau$) in nanoseconds is on the X-axis. Pink lines denote the direction $\mathrm{PC}_2 \rightarrow \mathrm{PC}_{3,4,5}$, whereas green lines denote the reverse direction. Shaded areas denote error bars computed over 20 independent realizations. The bottom panel shows the putative causal links summarizing the normalized IGT between \(\mathrm{PC}_2 \Leftrightarrow \mathrm{PC}_{3,4,5}\). Arrow lengths are proportional to the magnitude of the IGT, and the corresponding IGT values are reported next to each arrow. \(\mathrm{PC}_2\) exhibits bidirectional, symmetric causal links with \(\mathrm{PC}_{3,4,5}\).
}

\label{fig_SI:ntl9_pc2_graph}
\end{figure}

\begin{figure}[htbp]
    \centering
    \includegraphics[width=1.0\linewidth]{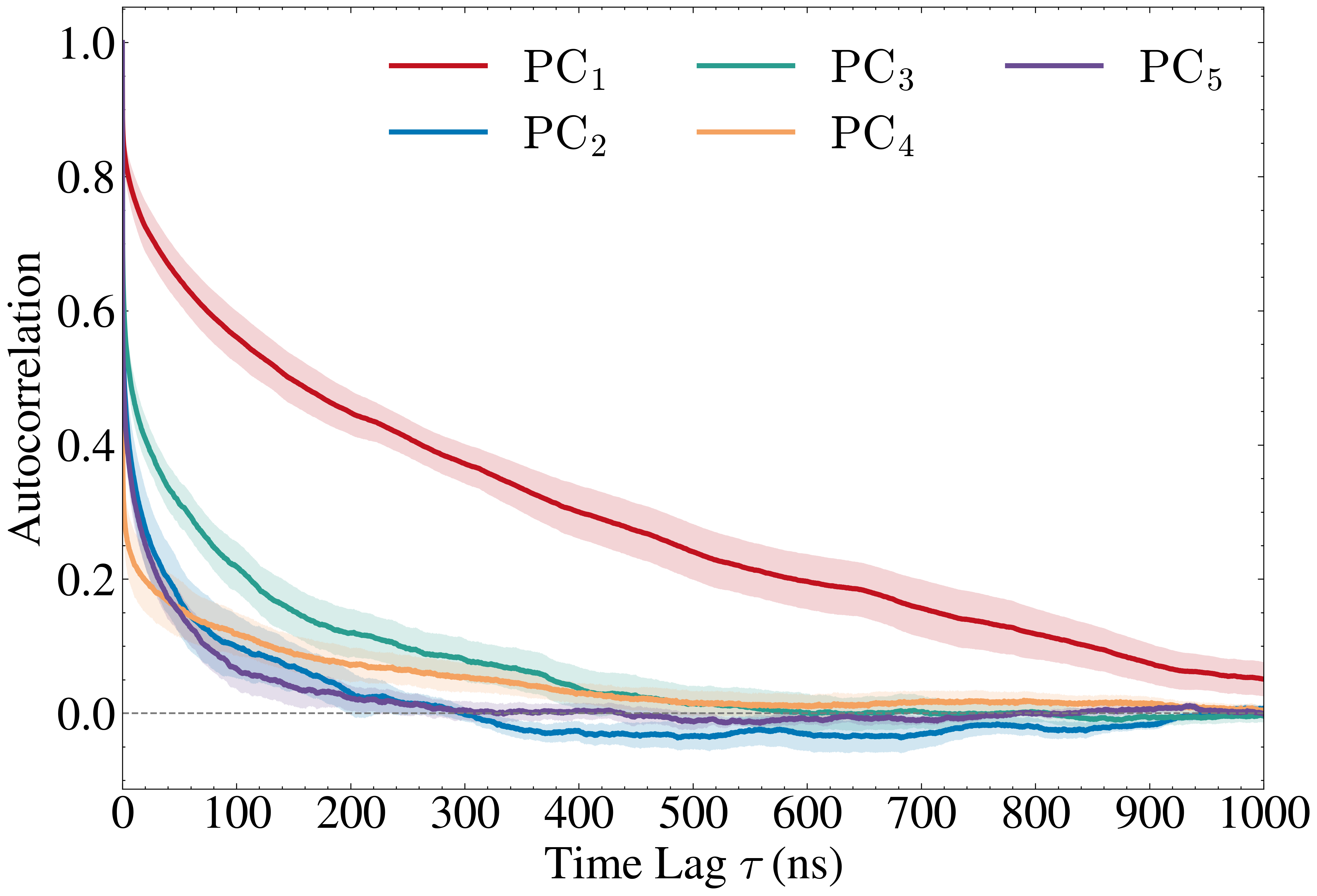}
    \caption{The autocorrelation functions of the top 5 PCs, $\mathrm{PC}_{i=1,5}$, which are colored red, blue, green, orange, and purple respectively, with shaded areas denoting error bars.}
    \label{fig_SI:ntl9-0_acf}
\end{figure}

\begin{figure}[htbp]
    \centering
    \includegraphics[width=1.0\linewidth]{figs/ntl9_0-pc1-ig-curves.png}
    \caption{The pairwise IG curves over time between $\mathrm{PC}_1 \Leftrightarrow \mathrm{PC}_{2,3,4,5}$. The IG values are on the Y-axis, and time lag ($\tau$) in nanoseconds is on the X-axis. Pink lines denote the direction $\mathrm{PC}_1 \rightarrow \mathrm{PC}_{2,3,4,5}$, whereas green lines denote the reverse direction. Shaded areas denote error bars computed over 20 independent realizations.}
    \label{fig_SI:ntl9-0_ig_curves}
\end{figure}

\begin{figure}[htbp]
\centering
\resizebox{1.0\linewidth}{!}{

\begin{tikzpicture}[
    PC/.style={circle, draw=black!70, fill=black!5, minimum size=1.8cm, font=\Large\sffamily, thick},
    PC1/.style={circle, draw=red!70!black, fill=red!20, minimum size=1.8cm, font=\Large\sffamily, thick}
]


    \node[PC1] (PC1) at (0,0) {PC$_1$};

    \def\R{7} 

    \node[PC] (PC2) at (195:\R) {PC$_2$};
    \node[PC] (PC3) at (245:\R) {PC$_3$};
    \node[PC] (PC4) at (295:\R) {PC$_4$};
    \node[PC] (PC5) at (345:\R) {PC$_5$};

    \begin{scope}[on background layer]
        \foreach \i in {2,3,4,5} {
            \draw[loosely dotted, very thick, purple!50!gray] (PC1) -- (PC\i);
        }
    \end{scope}

    \newcommand{\proparrow}[4]{
        \path (#1) -- (#2) coordinate[pos=0] (bStart) coordinate[pos=1] (bEnd);
        
        \pgfmathsetmacro{\vfrac}{#3 * 0.5}
        
        \draw[-{Latex[length=4.5mm, width=3.5mm]}, line width=2.5pt, draw=black!90]
            (bStart) -- ($ (bStart) ! \vfrac ! (bEnd) $)
            node[midway, #4, font=\normalsize\sffamily, inner sep=5pt] {#3};
    }

    \proparrow{PC1}{PC2}{1.00}{sloped, above}
    \proparrow{PC2}{PC1}{0.39}{sloped, below}

    \proparrow{PC1}{PC3}{0.67}{sloped, above}
    \proparrow{PC3}{PC1}{0.34}{sloped, below} 

    \proparrow{PC1}{PC4}{0.95}{sloped, above}
    \proparrow{PC4}{PC1}{0.66}{sloped, below}

    \proparrow{PC1}{PC5}{0.88}{sloped, above}
    \proparrow{PC5}{PC1}{0.41}{sloped, below}

\end{tikzpicture}
} 
\caption{Putative causal links between \(\mathrm{PC}_1\) and the remaining leading principal components. 
Nodes represent the first five PCs, and directed edges indicate the normalized integrated information transfer, \(\mathrm{IGT}\), defined in Eq.~\ref{eq:igt}. 
For each pair, the edge length, as well as the numerical label report the relative strength of the directional information transfer, with larger values corresponding to stronger causal links.}
\label{fig_SI:ntl9-0_pc1_graph}
\end{figure}

\begin{figure}[htbp]
  \centering
  \includegraphics[width=1.0\linewidth]{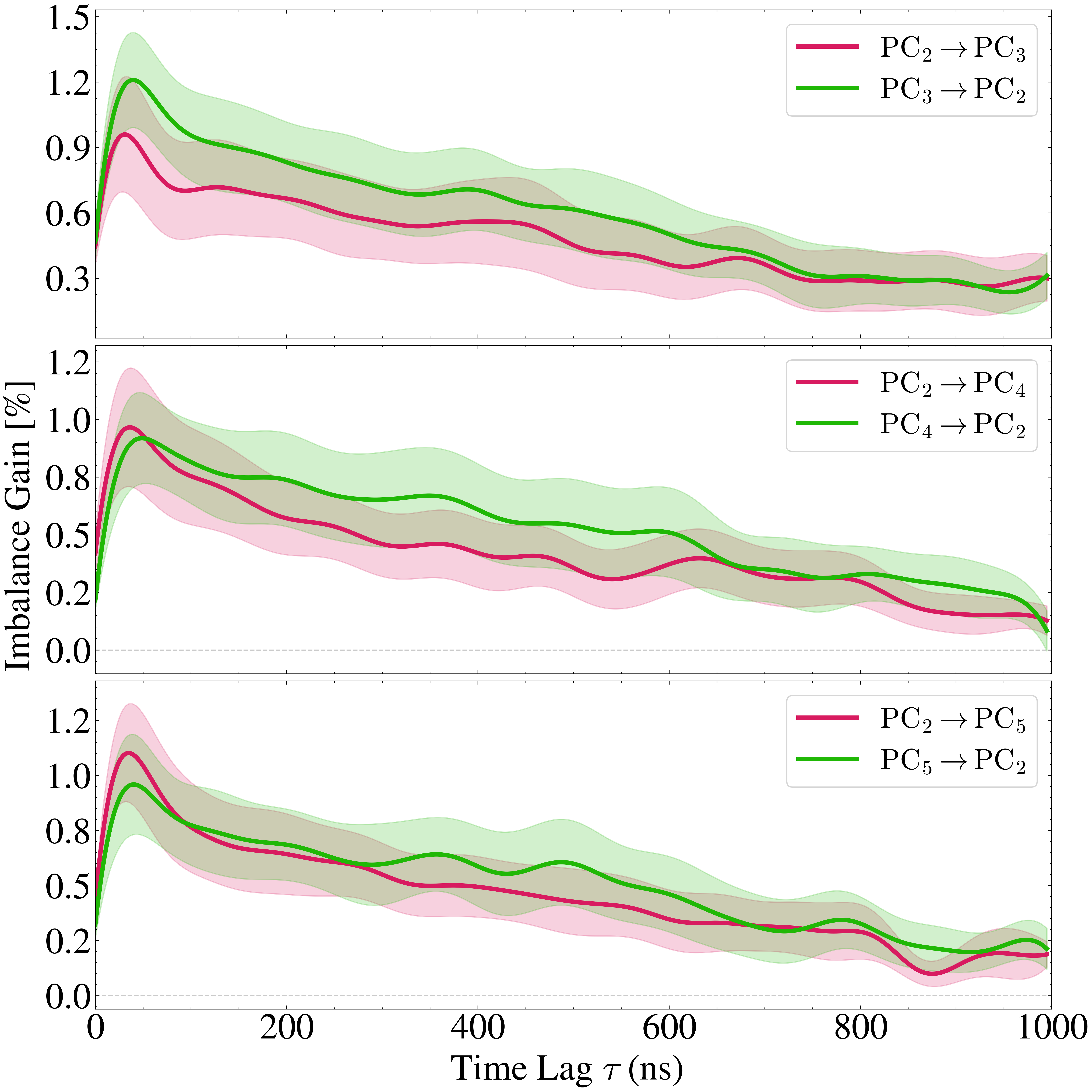}
\caption{The pairwise IG curves over time between $\mathrm{PC}_2 \Leftrightarrow \mathrm{PC}_{3,4,5}$. The IG values are on the Y-axis, and time lag ($\tau$) in nanoseconds is on the X-axis. Pink lines denote the direction $\mathrm{PC}_2 \rightarrow \mathrm{PC}_{3,4,5}$, whereas green lines denote the reverse direction. Shaded areas denote error bars computed over 20 independent realizations.
}
  \label{fig_SI:ntl9-0_pc2_ig_curves}
\end{figure}

\begin{figure}[htbp]
  \centering
  \includegraphics[width=1.0\linewidth]{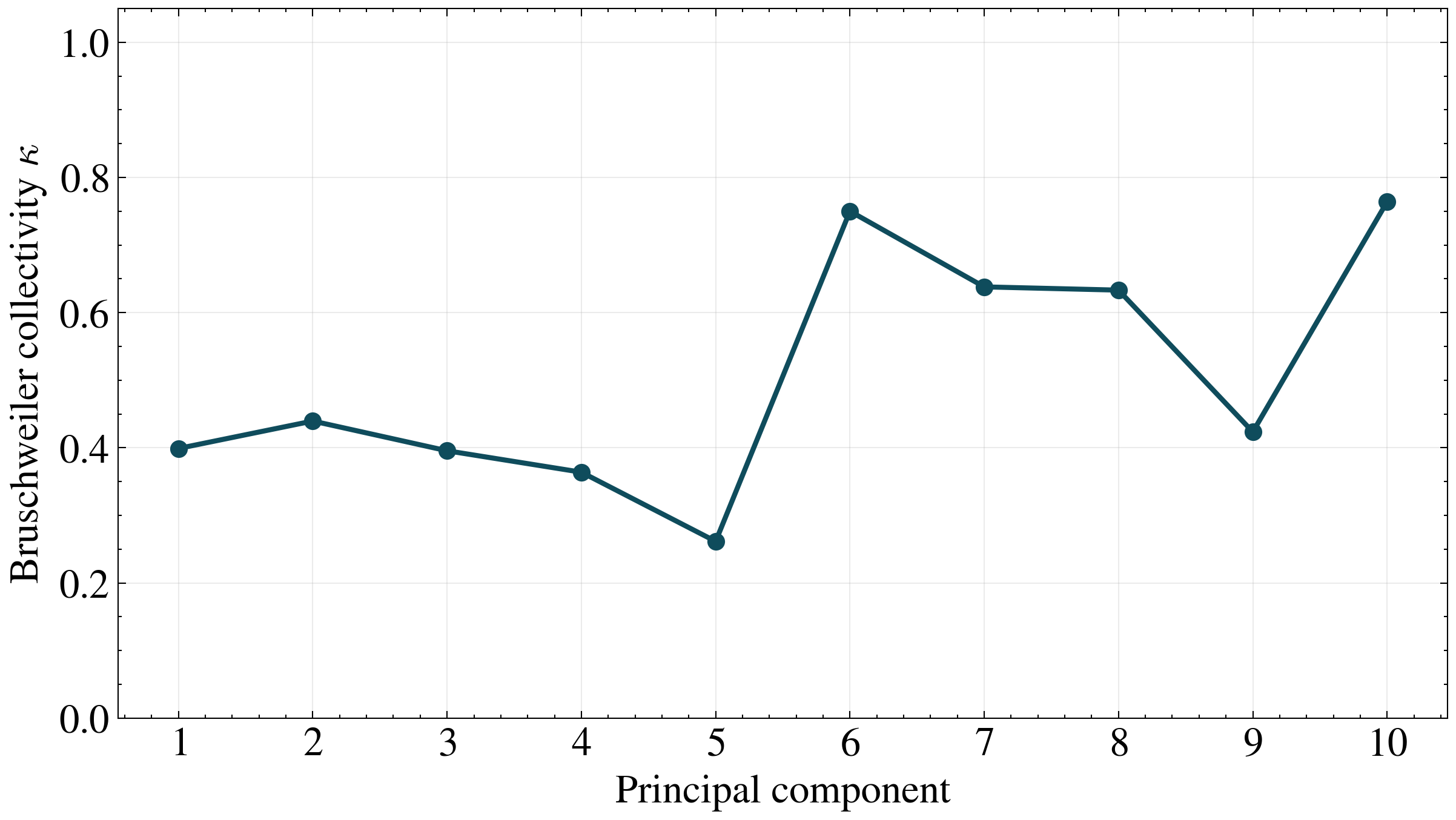}
\caption{Br\"uschweiler collectivity index for the leading PCs. The collectivity index quantifies the spatial extent of each mode.
}
  \label{fig_SI:ntl9-0_bruschweiller}
\end{figure}




\end{document}